\documentclass[twocolumn,nofootinbib,superscriptaddress]{revtex4-2}

\usepackage{graphicx}
\usepackage{dcolumn}
\usepackage{bm}

\usepackage{newtxtext,newtxmath}
\usepackage{amsmath}
 
\usepackage{amssymb}
\usepackage{color}
\usepackage{float}
\usepackage{subfigure}
\usepackage[T1]{fontenc}

\usepackage[colorlinks=true]{hyperref}
\hypersetup{citecolor = blue}

\newcommand{\SYSU}{School of Physics and Astronomy, Sun Yat-Sen University, Guangzhou 510297, P.R.China}
\newcommand{\PC}{Peng Cheng Laboratory, No. 2, Xingke 1st Street, Shenzhen 518000, P. R. China}
\newcommand{\SHJT}{Department of Astronomy, Shanghai Jiao Tong University, Shanghai 200240, P. R. China}

\begin{document}

\preprint{APS/123-QED}

\title[MCF for AP test]{Tomographic Alcock-Paczynski Test with Marked Correlation Functions}

\author{Liang Xiao}
\affiliation{\SYSU}
 
\author{Limin Lai}%
\affiliation{\SHJT}

\author{Zhujun Jiang}
\affiliation{\SYSU}

\author{Xiao-Dong Li}
\thanks{lixiaod25@mail.sysu.edu.cn}
\affiliation{\SYSU}

\author{Le Zhang}
\thanks{zhangle7@mail.sysu.edu.cn}
\affiliation{\SYSU}
\affiliation{\PC}




\date{\today}

\begin{abstract}
 The tomographic Alcock-Paczynski (AP) method,  developed over the past decade, exploits redshift evolution for cosmological determination, aiming to mitigate contamination from redshift distortions and capture nonlinear scale information. Marked Correlation Functions (MCFs) extend information beyond the two-point correlation. For the first time, this study integrated the tomographic AP test with MCFs to constrain the flat $w$CDM cosmology model. Our findings show that multiple density weights in MCFs outperform the traditional two-point correlation function, reducing the uncertainties of the matter density parameter $\Omega_m$ and dark energy equation of state $w$ by 48\% and 45\%, respectively. Furthermore, we introduce a novel principal component analysis (PCA) compression scheme that efficiently projects high-dimensional statistical measurements into a compact set of eigenmodes while preserving most of the cosmological information. This approach retains significantly more information than traditional coarse binning methods, which simply average adjacent bins in a lossy manner, yielding an additional $\sim 40\%$ reduction in error margins. To assess robustness, we incorporate realistic redshift errors expected in future spectroscopic surveys. While these errors modestly degrade cosmological constraints, our combined framework, which utilizes MCFs and PCA compression within tomographic AP tests, is less affected and always yields tight cosmological constraints. This scheme remains highly promising for upcoming slitless spectroscopic surveys, such as the Chinese Space Station Telescope (CSST).
\begin{description}
\item[Keywords]
large scale structure of universe, dark energy, cosmological parameters, marked weighted correlation function
\end{description}
\end{abstract}

\maketitle


\section{Introduction}

As of the present day, it remains an important problem to determine the nature of dark energy and dark matter. The large-scale structure (LSS) of galaxy clusters contains vast information about the expansion and structural growth history of our universe, which is still essential in unveiling the nature of dark energy and dark matter.

Over the past two decades, the Stage-III galaxy surveys, such as \citep{2df:Colless:2003wz,beutler20116df,blake2011wigglez,blake2011wigglezb,york2000sloan,Eisenstein:2005su,Percival:2007yw,
anderson2012clustering,alam2017clustering}, have expanded the boundaries of our understanding of the universe. Ongoing and upcoming in the next decade, advanced optical telescopes, such as the Dark Energy Spectroscopic Instrument (DESI)\footnote{https://desi.lbl.gov/},
the Vera C. Rubin Observatory (LSST)\footnote{https://www.lsst.org/},
the Euclid satellite\footnote{http://sci.esa.int/euclid/},
the Roman Space Telescope\footnote{https://roman.gsfc.nasa.gov/},
and the Chinese Space Station Telescope (CSST)\footnote{http://nao.cas.cn/csst/},  will conduct Stage-IV surveys. These surveys will accurately map the distribution of cosmic structures across a spatial scale of 10 billion light-years and characterize cosmic evolution over 10 billion years. Through statistical analysis of the upcoming survey data, we aim to understand the nature of dark matter and dark energy with a precision higher than 1\%. In particular, the precision of these surveys will allow us to explore non-linear clustering regions of galaxies, which contain a wealth of information. This poses a significant challenge for the cosmological analysis of the data. We need new statistical analysis methods that can effectively handle the nonlinear region, in order to comprehensively utilize the observational results and obtain more physical information.

The Alcock-Paczynski (AP) test~\citep{AP1979} serves as a means to probe the expansion history of the Universe. In essence, the test can be applied to any structure in the Universe displaying random orientations, such as halos, filaments, or walls, leading to a spherically symmetric ensemble average. Under a specific cosmological model, the radial and tangential sizes of distant objects or structures can be expressed as $\Delta r_\parallel$ and $\Delta r_{\perp}$ (Eq.\ref{eq:r}). If the models deviate from the true cosmology, the computation of $D_A$ and $H(z)$ results in an incorrect estimation of $\Delta r_ \parallel$ and $\Delta r_\perp$. This inaccuracy manifests as geometric distortions along the line of sight (LOS) and perpendicular directions, referred to as the AP distortions. The AP distortions can be measured and quantified through statistical analysis of the large-scale galaxy distribution. Consequently, the observed distortions offer a method to constrain cosmological parameters.

However, the expansion history can only be probed with structures that have not decoupled from the Hubble flow through gravitational collapse. Moreover, peculiar LOS motions of observed objects within the structure lead to a deviation from spherical symmetry. Thus, to apply the AP test, one must address these redshift-space distortions (RSD), requiring the modeling of peculiar velocities.

Recently, \citep{LI14, LI15} introduced the "tomographic AP method" as a notable advancement, highlighting its ability to effectively mitigate the impact of RSD contamination. The key concept of the method utilizes the redshift evolution of LSS anisotropy, which is sensitive to the AP effect, while remaining insensitive to distortions caused by RSD. This enables the separation of AP distortion from RSD contamination. In the literature, \citep{LI14, LI15, Park:2019mvn} applied the AP test to LSS, yielding precise constraints on cosmological parameters related to cosmic expansion. \citep{LI15} introduced a novel approach by quantifying anisotropic clustering through $\widehat{\xi}_{\Delta s}(\mu)$ (Eq.\ref{eq:xi_normalization}), defined as the integration of the two-dimensioanl two-point correlation functions (2PCF), $\xi(s, \mu)$, over the clustering scale $s$ and normalization alone with $\mu$. Subsequently, \citep{LI16} pioneered the application of this method to the Sloan Digital Sky Survey (SDSS), Baryon Oscillation Spectroscopic Survey (BOSS), and DR12 galaxies, resulting in a remarkable 35\% enhancement in constraints on the allowed parameter space $\Omega_m$--$w$ for dark matter and dark energy equation of state. Further investigations by \citep{LI18} and \citep{Zhang2019} demonstrated the substantial improvement in constraints on dynamical dark energy models. Additionally, \citep{LI19} analysis revealed that when combining the method with Cosmic Microwave Background (CMB) and Baryon Acoustic Oscillation (BAO) datasets, using DESI-like data, the dark energy figure of merit \citep{Wang_2008} could be enhanced by an order of magnitude, while \citep{Luo_2019} demonstrated the feasibility of conducting tomographic AP analysis in Fourier space. In 2023, \citep{Dong_2023} employed the methodology developed by \citep{Park:2019mvn} to analyze SDSS galaxy data, deriving a constraint on the dark energy equation of state parameter of $w=0.903\pm0.023$ when combined with SNIa and BAO measurements. This result exhibits a 4.2$\sigma$ tension with the predictions of the $\Lambda$CDM model. Notably, it aligns with the later findings from DESI \citep{DESI:2024mwx,DESI:2025zgx,DESI:2025fii}, which also favor $w>-1$ at low redshifts.

A key limitation of the current tomographic AP method is its exclusive reliance on two-point statistics. While this approach is well-established, two-point statistics may be insufficient for extracting cosmological information from non-Gaussian fields. This limitation is particularly relevant for upcoming stage-IV surveys, which will probe the non-linear clustering regime where the density field becomes highly non-Gaussian. To enhance its constraining power, a logical extension would be to apply the AP method to statistics beyond 2PCF and power spectrum. A recent addition to this field is the density-marked correlation functions (MCFs)\citep{stoyan_correlations_1984}, offering a density-dependent statistic that extends beyond two-point statistics and can effectively capture non-Gaussian features~\citep{Beisbart2000,Beisbart2002,Gottl2002,Sheth:2004vb,Sheth:2005aj,Skibba2006,White_2009,White_2016,Satpathy:2019nvo,2020arXiv200111024M,PMS2020,Massara:2022zrf,Jung:2024esv,Philcox:2020fqx, Philcox:2020srd, Ebina:2024zkv, Massara:2024cvu}. The MCFs introduce a weight for each galaxy using a ``marker'' based on its local density, thereby providing density-dependent clustering information. The conventional 2PCF approach is, of course, a specific case of MCFs, where all galaxies receive identical weights irrespective of differences in their physical properties and environments. The distinctive feature of MCFs lies in its ability to highlight statistics in regions of higher or lower density. We have the freedom to choose weights that allow the statistic to respond to both positive and negative powers of the density, making MCFs highly versatile. More recently, studies using N-body simulations have demonstrated that combining different weights in MCFs can significantly improve cosmological parameter constraints \citep{Yang:2020ysv, XiaoYuan_2022}. Based on these findings, \citep{Lai:2023dzp, Yin:2024iqp} applied MCFs to SDSS galaxy data, further confirming enhanced parameter constraints compared to traditional methods.

Another growing challenge in modern surveys is the increasingly complexity of covariance matrix estimation. This is mainly due to two aspects: 1) as surveys advance we must use more complex statistics, also for the wider redshift coverage of the stage-IV surveys we need more redshift bins, this both increases data vector sizes and makes covariance calculations much harder; 2) larger volume and higher precision survey samples require larger and higher resolution samples for covariance matrix, making simulation samples preparation particularly difficult and expensive. In this study, the combined MCFs also leads to significant expansion of size of statistics. To solve this problem, we adopt a compression scheme based on principal component analysis (PCA)~\citep{Huterer_2003, Crittenden_2009} to address this issue. PCA is an efficient technique to reduce the size of statistics of combining different types of MCFs in this study. Besides, using PCA we can also compress the full $(s,\mu)$-dependence of the correlation functions, which is extremely difficult to utilize in the previous cosmological studies work due to the extremely difficult covariance estimation.

The redshift uncertainties in slitless spectroscopy may also impact the performance of the tomographic AP test. Tomographic AP test is sensitive to the intrinsic anisotropy in the survey data. In the next few years, slitless instruments will be used in CSST~\citep{Gong_2019} and the Euclid satellite~\citep{EUCLID} to determine the spectroscopic redshifts of galaxies. While slitless spectroscopy enhances the efficiency and practicality of spectroscopic surveys, it introduces significantly larger redshift errors compared to conventional fiber-based methods. Our analysis in ~\cite{Xiao_2022} shows the tomographic AP test's constraining power experiences mild degradation under conservative CSST assumptions ($\sigma_z \lesssim 0.002$). In this study, we further test the impact of redshift errors caused by slitless instruments in the cases of MCF, and the cases of MCFs and ($s,\mu$)-dependency 2-dimensional MCFs with PCA compression. Following~\cite{Xiao_2022}, in this study, we will not only test the impact of the conservative estimate for CSST, but also test a larger estimate and an estimate with a deviation of linear between errors and redshifts.

In this study, utilizing a large N-body simulation, we will, for the first time, assess the performance of MCFs in constraining cosmological parameters, with a focus on $\Omega_m$ and $w$, using the tomographic AP test. The paper is structured as follows: In Sect.~\ref{sec:data}, we introduce the simulation data used for our analysis. Sect.~\ref{sec:methods} presents the methodology for calculating the MCF on mock data and the tomographic AP test. In Sect.~\ref{sec:results}, we present our findings and compare the constraints between the standard 2PCF and MCFs. We also highlight constraints improved by using the combined MCFs with different weights and the PCA compression method. As one of the follow-up works of~\cite{Xiao_2022}, the impact of redshift error from slitless instruments on combined MCFs and PCA compression is also considered. Besides, a robustness check about the systematics estimation is attached to clarify the current limitations of our method. Finally, we summarize our results and conclude in Sect.~\ref{sec:con}.

\section{Data}\label{sec:data}

We employ the Big MultiDark (BigMD) simulation \citep{BIGMD}, one of the largest within the MultiDark series, as our mock data for the analysis of the tomographic AP method. It comprises $3840^3$ particles in a cubic box with a side length of $(2.5h^{-1}~{\rm Gpc})^3$. The simulation adopts a $\Lambda$CDM cosmology model consistent with Planck observations, with parameters set to $\Omega_m = 0.307115$, $\Omega_b = 0.048206$, $\sigma_8 = 0.8288$, $n_s = 0.9611$, and $h = 0.6777$ \citep{BIGMD}. The initial conditions for the simulation were established using the Zeldovich approximation \citep{Crocce2LPT} at a redshift of $z_{\rm init} = 100$. This results in a mass resolution of about $2.4\times 10^{10}$ $h^{-1} M_{\odot}$. 

Throughout this analysis, we use the halos and subhalos identified by the ROCKSTAR halo finder \citep{ROCKSTAR} at snapshots $z_1=0.607$ and $z_2=1$, resulting in halo catalogs comprising  snapshots generated from the simulated particles. It is important to note that when dealing with real observational data, it becomes essential to incorporate a lightcone approach to model systematics effectively \citep{LI16}. However, in this study, where our primary concern is the impact of redshift errors arising from RSD, we opt for simplicity and utilize snapshots.

We investigate the variation in the anisotropic correlation function measured at two snapshots to reveal the redshift evolution of anisotropic clustering. To maintain a consistent number density, we enforce $\bar{n} = 0.001$ $(h^{-1}\rm Mpc)^{-3}$ by implementing a mass cut on halos and subhalos. Specifically, it is:

\begin{equation}
    m_H \in 
    \begin{cases}
    [3.1\times10^{12}, 1.1 \times 10^{15}] M_{\odot}/h, &  z = 1.0 \\ 
    [3.9 \times 10^{12}, 2.2 \times 10^{15}] M_{\odot}/h, & z = 0.6069 \\ 
    \end{cases}\label{eq:old_mass_threshold}
\end{equation}

\noindent This density is comparable to the galaxy number density expected in current and next-generation spectroscopic galaxy surveys. The substantial volume, high resolution, and broad redshift coverage sufficiently meet our testing requirements.

\section{Methods}\label{sec:methods}

The AP effect denotes the observed geometric distortions in LSS that occur when incorrect cosmological models are applied to convert galaxy redshifts into comoving distances. For a distant object or structure in the universe, one can calculate its size along and perpendicular to LOS using the relations provided by
\begin{equation}
\Delta r_{\parallel} = c \frac{\Delta z}{H(z)}, \  
\Delta r_{\perp} = (1+z) D_A(z) \Delta \theta\,, \label{eq:r}
\end{equation}
where $\Delta z$ and $\Delta \theta$ represent the observed redshift span and angular size measured through observations, and $H(z)$ and $D_A(z)$ stand for the Hubble parameter and the angular diameter distance, respectively. 

We adopt the spatially-flat $w$CDM cosmology, characterized by two free parameters: the matter density parameter at the present day, $\Omega_m$, and the constant dark energy equation of state ($w$). In this model, the redshift-dependent functions $H(z)$ and $D_A(z)$ can be expressed by 
\begin{equation}
    \begin{aligned}
    H(z) &= H_0 \sqrt{\Omega_m(1+z)^3 + (1-\Omega_m)(1+z)^{3(1+w)}}, \\ 
        D_A(z) &= \frac{c}{1+z} \int_0^z \frac{\mathrm{d} z'}{H(z')}\,,
    \end{aligned}
\end{equation}\label{eq:H_and_D}
where $H_0$ denotes the current Hubble parameter. Therefore, adopting an incorrect combination of $\Omega_m$ and $w$ would lead to misestimations of $\Delta r_\parallel$ and $\Delta r_\perp$. Consequently, the resulting objects exhibit distorted shapes, reflecting the presence of the AP effect. In galaxy surveys, measuring galaxy clustering in both radial and transverse directions allows for the determination of $D_A$ and $H(z)$, providing a means to constrain the relevant cosmological parameters.

\subsection{Tomographic AP test}\label{sec:AP_2PCF}

Ref. \cite{LI14}  introduced a tomographic analysis of small-scale galaxy clustering aimed at effectively distinguishing the AP effect from RSD. They employed the integrated 2-dimensional 2PCF along various LOS to evaluate the level of anisotropy across different redshifts. This is represented by 
\begin{equation}\label{eq:xi}
    \xi_{\Delta s}(\mu) = \int_{s_{\rm min}}^{s_{\rm  max}} \xi(s, \mu) ds\,,
\end{equation}
where $s$ represents the comoving distance between pairs and $\mu$ is defined as $\cos(\theta)$, with $\theta$ being the angle between the LOS direction and the line connecting the pair. The integral limits are specified as $s_{\rm min}= 6~h^{-1}{\rm Mpc}$ and $s_{\rm max} = 40~h^{-1}{\rm Mpc}$. Following our previous investigations, in short, choosing this integration range for $s$ is optimal for measuring the AP effect and providing robust cosmological constraints.

We adopt the widely used Landy–Szalay estimator \citep{Landy} to calculate the 2PCF
\begin{equation}\label{eq:xi_primary}
    \xi(s, \mu) = \frac{DD-2DR+RR}{RR}\,,
\end{equation}
where $DD$ represents the number of galaxy-galaxy pairs, $DR$ corresponds to the galaxy-random pairs, and $RR$ denotes the random-random pairs. 

To address the systematic uncertainty arising from galaxy bias and clustering strength, an additional normalization is enforced by
\begin{equation}\label{eq:xi_normalization}
    \widehat{\xi}_{\Delta s}(\mu) = \frac{\xi_{\Delta s}(\mu)}{\int_{0}^{\mu_{\rm max}}\xi_{\Delta s}(\mu)}\,,
\end{equation}
where we apply a cutoff $\mu < \mu_{\rm max}$ to mitigate the non-linear Finger-of-God (FoG) effect and severe fiber collisions, which become particularly significant along LOS as $\mu$ approaches 1.

The primary objective of the tomographic AP method is to eliminate the RSD effect. The essence of this method is to use the difference in the normalized anisotropy $\widehat{\xi}(\Delta s, \mu)$ between two redshifts to evaluate the redshift evolution of LSS distortion. This contrast is primarily influenced by the AP distortion, displaying significant insensitivity to the RSD distortion. This characteristic allows us to avoid substantial contamination from RSD and explore information about AP distortion on modest clustering scales. Specifically, the difference between the $i$-th and $j$-th redshift bins is expressed as 
\begin{equation}\label{eq:deltaxi}
    \delta \widehat{\xi}_{\Delta s}(z_i, z_j, \mu) = \widehat{\xi}_{\Delta s}(z_i, \mu) -\widehat{\xi}_{\Delta s}(z_j, \mu)\,.
\end{equation}
For real observations, it is beneficial to adopt multiple pairs of redshift slices to enhance constraining power. While various redshift slicing approaches are feasible \citep{LI16, LI19}, the specific selection employed in this study, with $z_1 = 1.0$ and $ z_2 = 0.6069$, is indicative of stage-IV surveys that typically obtain numerous spectroscopic galaxy samples around $z \sim 1$. For CSST, the slitless redshift errors may become large at $z \gtrsim 1$ due to its potential inability to detect Ly-$\alpha$ lines in galaxies with redshifts greater than 1. Additionally, \cite{Ma_2020} demonstrated that the systematics of the tomographic AP method are notably larger for redshift slices $z \gtrsim 1$. Therefore, we opt not to include additional tomography slices at higher redshifts.

Additionally, effects like the RSD contribute to a non-zero redshift evolution of anisotropy in $\delta \widehat{\xi}_{\Delta s}$. Furthermore, apart from the RSD effect, various sources of systematics originate from observational effects like selection bias and fiber collision. To avoid the complexities associated with modeling these systematics, a practical approach involves correcting for them using mock surveys \citep{LI14, LI15, LI16}. In this context, $\delta \widehat{\xi}_{\Delta s}^{\rm corr}(\mu)$ represents the redshift dependence after correcting for systematics for the selected two redshift bins, defined as
\begin{equation}
  \delta \widehat{\xi}_{\Delta s}^{\rm corr}(\mu)    = \delta \widehat{\xi}_{\Delta s}(z_1, z_2, \mu) - \delta \widehat{\xi}_{\Delta s, {\rm sys}}(z_1, z_2, \mu) \label{eq:correlation_xi}.
\end{equation}
Here, the correction term $\delta\widehat{\xi}_{\Delta s, {\rm sys}}$ accounts for RSD-dominated systematic effects, which induce a non-zero redshift evolution of anisotropy, and is estimated by mock in the fiducial cosmology, both in the tests of simulation level and observation level. This correction has to be estimated only by mock rather than perturbation theories because the clustering scale is small: 6-40$\mathrm{Mpc}/h$ (See Appendix \ref{sec:appendix_RSD} for details). In addition, we found it to be largely insensitive to the choice of cosmological models, as verified through tests \citep{LI15, Park:2019mvn} (besides, Appendix \ref{sec:appendix_RSD}), and is therefore determined using the \texttt{BigMD} mock surveys. Therefore, it is an unbiased estimation in this work. 

We have adopted binning for $\mu$, specifically using 17 coarse $\mu$-bins with equal width in $0.0 < \mu < 0.97$ (i.e. $r_{\mathrm{max}} = 0.97$ in Eq.\ref{eq:xi_normalization}), which is averaged from 116 narrow bins and has removed the last element to eliminate the correlation caused by averaging. We cut $\mu > 0.97$ to weaken the impacts of FOG. 

In Bayesian inference of model parameters ($\theta$), we aim to evaluate the posterior probability distribution
\begin{equation}
p(\theta |d)\propto  \mathcal{L}(d|\theta) p(\theta)\,,
\end{equation}
where $ p(\theta)$ denotes a prior probability distribution of $\theta$.  This study focuses on constraining two cosmological parameters: $\Omega_m$ and $w$. We adopt flat priors for these parameters, in the range of $\Omega_m \in [0.2, 0.5]$ and $w \in [-1.5,0.5]$.  

The term $\mathcal{L}(d|\theta)$ stands for the likelihood, which is defined as    
\begin{equation}\label{eq:likeli}
-2\ln \mathcal{L}(d|\theta)= \chi^2\,,
\end{equation}
with

\begin{equation}\label{eq:chi2}
\chi^2 =  \big(\boldsymbol{p}(\theta)- \boldsymbol{p}_{\rm true} \big)^T  \boldsymbol{C}^{-1} \big(\boldsymbol{p}(\theta)-\boldsymbol{p}_{\rm true} \big)\,.
\end{equation}
Here, the vector $\boldsymbol{p} \equiv \delta \widehat{\xi}_{\Delta s}^{\rm corr}(\mu)$ (as defined in Eq.~\ref{eq:correlation_xi}) represents the redshift evolution of the anisotropy caused by the AP effect, which is dependent on the cosmological parameters. Specifically, $\boldsymbol{p}=[\delta \widehat{\xi}_{\Delta s}^{\rm corr}(\mu_1), \ldots, \delta \widehat{\xi}_{\Delta s}^{\rm corr}(\mu_n)]$ collects all variables $\delta \widehat{\xi}_{\Delta s}^{\rm corr}(\mu_i)$ for all $\mu$-bins. Given the number of $\mu$-bins, there are a total of 17 variables in $\boldsymbol{p}$. Thus, $\chi^2$ quantifies the degree to which anisotropy deviates from zero, where a value of zero corresponds to an isotropic true cosmological background after the RSD correction.

In practice, the true cosmological parameter values are unknown. Thus, determining the parameter $\theta$ corresponds to minimizing the tomographic AP effect, which is quantified through the redshift evolution of anisotropy.  To compute the redshift evolution of anisotropy, we must assume a fiducial geometric background. We adopt a cosmology with $\Omega_m = 0.3071$ and $w = -1.0$ as the fiducial model. Notably, the parameter estimation is largely insensitive to the choice of the fiducial background.

Following~\citep{Xiao_2022}, the covariance matrix $\boldsymbol{C}$ is estimated from the $\mathtt{BIGMD}$ simulation. We initially divide the entire simulation box into $8^3$ sub-boxes, each with a side length of 325 $h^{-1}{\rm Mpc}$. Subsequently, we employ empirical covariance matrix estimation based on these 512 sub-boxes.  The number of sub-box samples is larger than the size of the $\boldsymbol{p}$, which ensures the convergence of the covariance matirx. We do a robustness check of covariance matrix in Appendix \ref{sec:appendix_cov}. The statistical uncertainty of the correction term is also negligible compared to $\delta \widehat{\xi}_{\Delta s}(z_1, z_2, \mu)$ (see Eq.~\ref{eq:correlation_xi}). Therefore, the covariance only accounts for the contribution from the fluctuations of $\delta \widehat{\xi}_{\Delta s}(z_1, z_2, \mu)$. Note that we choose sub-boxes located at different spatial positions to compute $\delta \widehat{\xi}_{\Delta s}(z_1, z_2, \mu)$. In this case, the comparison between two redshfit bins is always made for sub-samples with different cosmic variance, so that to avoid the effect of related noise fluctuations and structural features between the two snapshots.

\subsection{MCF statistics}

In the following, we will incorporate MCFs into the tomographic AP test, and investigate in detail the performance of MCFs and its impact on the improvement of cosmological parameter constraints, comparing it with the result from 2PCF. Let us begin by providing a brief description of the analysis framework of MCFs.

MCFs extend the standard 2PCF, a statistical measure assigning weights to objects before measuring their correlation \citep{White_2016, Yang:2020ysv, XiaoYuan_2022}. Physically, we assign weights based on their local densities to consider environmental dependence in our analysis. MCFs are sensitive to the chosen weight, as clustering and redshift space distortion effects exhibit notable variations in dense and sparse regions of the universe. As outlined in \cite{Yang:2020ysv}, these weights are designed in the following way
\begin{equation}
    \label{eq:weightrho}
w(\boldsymbol{x})=\rho_{n_{\mathrm{NB}}}^\alpha \,.
\end{equation}
Here the local density $\rho_{n_{\rm NB}}$ is calculated by considering its $n_{\rm NB}$ nearest neighbors,
\begin{equation}
    \rho_{n_{\mathrm{NB}}}(\boldsymbol{x})=\sum_{i=1}^{n_{\mathrm{NB}}} W_k\left(\boldsymbol{x}-\boldsymbol{r}_i, h_W\right)\,,
\end{equation}
where $\rho_{n_{\rm NB}}(\boldsymbol{x})$ is the environmental number density around a specific galaxy located at $\boldsymbol{x}$, and $W_k$ is the smoothing kernel, for which we choose the third-order B-spline functions which has non-zero values within a sphere of radius $h_W$ $h^{-1} {\rm Mpc}$. In this study, we set $n_{\rm NB}=30$, indicating that the radius $h_w$ is determined by the distance between a selected halo and its 30-th nearest neighbor~\citep{Gingold1977,Lucy1977}. 

Except for the weight assigned to each halo, the computational procedure for measuring the MCFs closely follows that used for the standard 2PCF. Similar to the computation of the traditional 2PCF through  $\xi(\boldsymbol{r})=\langle\delta({\bf x}) \delta(\boldsymbol{x}+\boldsymbol{r})\rangle$, the MCFs statistic is defined as  
\begin{eqnarray}
W^{\alpha}(\boldsymbol{r})=\left\langle\delta(\boldsymbol{x}) \rho^\alpha_{n_{\mathrm{NB}}}(\boldsymbol{x})\, \delta(\boldsymbol{x}+\boldsymbol{r}) \rho^\alpha_{n_{\mathrm{NB}}}(\boldsymbol{x}+\boldsymbol{r})\right\rangle\,.
\end{eqnarray}
Note the distinction between $\delta (\boldsymbol{x})$ and $\rho_{n_{\mathrm{NB}}}(\boldsymbol{x})$, where the former denotes the point-like density contrast, while the latter represents the smoothed density field. Certainly, setting $\alpha=0$ causes the MCFs to revert to the standard 2PCF.

The Landy–Szalay estimator in Eq.~\ref{eq:xi_primary} for MCFs is now expressed as 
\begin{equation}
    W^{\alpha}(s,\mu) = \frac{WW-2WR+RR}{RR}\,,
\end{equation}
where $WW$, $WR$ and $RR$ denote the weighted number of galaxy-galaxy pairs, the galaxy-random pairs, and the random-random pairs, respectively.  

Extending Eq.~\ref{eq:deltaxi} to the case of MCFs, we express the difference between the redshift bins $z_1$ and $z_2$ as
\begin{equation}\label{eq:hat_W_1d}
    \delta \widehat{W}_{\Delta s}^{\alpha}(\mu_i) = \widehat{W}_{\Delta s}^{\alpha}(z_1,\mu_i) - \widehat{W}_{\Delta s}^{\alpha}(z_2,\mu_i)\,,
\end{equation}
where we have dropped out $z_1$ and $z_2$ on the left-hand side for convenience. The normalized quantity $\widehat{W}_{\Delta s}^{\alpha}$ is calculated in the same manner as in Eq.~\ref{eq:xi_normalization}, namely
\begin{equation}\label{eq:w_normalization}
    \widehat{W}^{\alpha}_{\Delta s}(\mu) = \frac{W^{\alpha}_{\Delta s}(\mu)}{\int_{0}^{\mu_{\rm max}}W^{\alpha}_{\Delta s}(\mu)}\,,
\end{equation}  
where 
\begin{equation}\label{eq:w}
W^{\alpha}_{\Delta s}(\mu) = \int_{s_{\rm min}}^{s_{\rm  max}} W^{\alpha}(s, \mu) ds\,, 
\end{equation}
with the integration range identical to that in Eq.~\ref{eq:xi}, namely $s_{\rm min}= 6~h^{-1}{\rm Mpc}$ and $s_{\rm max} = 40~h^{-1}{\rm Mpc}$. 

The systematic correction of MCFs is similar to that of 2PCF: 
\begin{equation}
  \delta \widehat{W}_{\Delta s}^{\alpha, \rm corr}(\mu)    = \delta \widehat{W}^{\alpha}_{\Delta s}(z_1, z_2, \mu) - \delta \widehat{W}^{\alpha}_{\Delta s, {\rm sys}}(z_1, z_2, \mu) \label{eq:correction_W}.
\end{equation}

The 2PCF and MCFs of the halo catalogs within backgrounds of wrong cosmologies are obtained by applying the mapping approach using coordinate transformations in the level of 2PCF and MCFs instead of mock data, as detailed in Appendix A(especially, Eq. A1 and Fig. 6) of \citep{LI18}:
\begin{equation}\label{eq:coordinate}
\begin{gathered}
s=s_{\rm fid} \sqrt{\kappa_{\|}^2 \mu_{\rm fid }^2+\kappa_{\perp}^2\left(1-\mu_{\perp}^2\right)}\,, \\
\mu=\mu_{\rm fid} \frac{\kappa_{\|}}{\sqrt{\kappa_{\|}^2 \mu_{\rm fid }^2+\kappa_{\perp}^2\left(1-\mu_{\perp}^2\right)}}\,.
\end{gathered}
\end{equation}

\noindent where $\kappa_{\|} \equiv D_{A, \rm target} / D_{A,\rm fid}$ and $\kappa_{\perp} = H_{\rm fid} / H_{\rm target}$ and they are computed in the two snapshots of the target and fiducial cosmologies, fixing $\mathrm{Mpc}/h$ unit, respectively.  This approach is much more efficient than converting the halo samples into different backgrounds and remeasuring the 2PCFs and MCFs. However, it should be noted that: i) changes in the values of the weights are not captured by Eq.~\ref{eq:coordinate}. This discrepancy arises because the AP effect nonuniformly distorts the geometry, causing the set of $n_{\rm NB}$ nearest neighbors to differ from one cosmology to another; ii) only the background evolution of different cosmologies is considered, ignoring the effects arising from structure growth.
However, the former has been shown in previous studies~\citep{Yang:2020ysv} to have only a very minor impact, which can be safely ignored. The latter is not to be considered in the tomographic AP test, as the AP effect is the geometric distortion corresponding to the background evolution itself and we only utilize the background evolution of different cosmologies to constrain cosmology. 

When constraining cosmological parameters, \cite{Yang:2020ysv} has been observed that the constraining power of MCFs combining multiple different $\alpha$ values is significantly stronger than the results obtained using a single $\alpha$. Due to this observation, we will explore the combined MCFs with different $\alpha$, forming $\boldsymbol{p}^T=[\boldsymbol{p}_{\alpha_1}^{T},\boldsymbol{p}_{\alpha_2}^{T},\cdots,\boldsymbol{p}_{\alpha_n}^{T}]$, with check the ordering in covariance
\begin{equation}
    \boldsymbol{p}_{\alpha_i} = \delta \widehat{W}_{\Delta s}^{\alpha_i}(\mu) - \delta \widehat{W}_{\Delta s, \rm sys}^{\alpha_i}(\mu)\,. \label{eq:p}
\end{equation}


We totally have 18 $\mu$-bins for each MCF, resulting in a total of 54 elements in the vector $\boldsymbol{p}$ for combined MCFs with three $\alpha$s as an example. For convenience, we mark the statistics in this case as $\delta W_{\Delta s}^{\boldsymbol{\alpha}}$ or $\delta W_{\Delta s}^{\boldsymbol{\alpha}, \rm corr}$ (with systematic correction) with specifying $\boldsymbol{\alpha} = [\alpha_1, \alpha_2, \cdots, \alpha_n]$, or combined MCFs in the following text. The cosmological parameters $\Omega_m$ and $w$ are constrained using the \texttt{emcee} Python package~\citep{emcee}, which implements a Markov Chain Monte Carlo (MCMC) sampling algorithm.

In this work, the full 2-dimensional 2PCF and MCFs are also applied to constrain cosmology parameters with PCA (see Sec.\ref{sec:method_PCA} for details). The full 2-dimensional 2PCF is defined as \citep{Park:2019mvn} 
\begin{equation}\label{eq:xi_2d_normalization}
    \widehat{\xi}_{\rm 2d}(s, \mu, z) = \frac{\xi_{\rm 2d}(s, \mu, z)}{2 \pi \int_{0}^{1} \rm d \mu \int_{s_{\rm min}}^{s_{\rm max}} \xi_{\rm 2d}(s, \mu,z)}\,,
\end{equation}
\noindent while the full 2-dimensional MCFs are defined as 
\begin{equation}\label{eq:w_2d_normalization}
    \widehat{W}_{\rm 2d}^{\alpha}(s, \mu, z) = \frac{W_{\rm 2d}^{\alpha}(s, \mu, z)}{2 \pi \int_{0}^{1} \rm d \mu \int_{s_{\rm min}}^{s_{\rm max}} W_{\rm 2d}^{\alpha}(s, \mu,z)}\,,
\end{equation}
\noindent respectively. Correspondingly, the systematic correction of full 2-dimensional 2PCF is 
\begin{equation}
  \delta \widehat{\xi}_{\rm 2d}^{\rm corr}(s, \mu)    = \delta \widehat{\xi}_{\rm 2d}(z_1, z_2, s, \mu) - \delta \widehat{\xi}_{\rm 2d, {\rm sys}}(z_1, z_2, s, \mu) \label{eq:correlation_xi_2d}, 
\end{equation}
\noindent where $\delta \widehat{\xi}_{2d}(z_1, z_2, s, \mu) = \delta \widehat{\xi}_{\rm 2d}(z_1, s, \mu) - \delta \widehat{\xi}_{\rm 2d}(z_2, s, \mu)$ and $\delta \widehat{\xi}_{\rm 2d,\rm sys}(z_1, z_2, s, \mu)$ is calculated in the fiducial cosmology. The systematic correction of full 2-dimensional MCFs are similiar to that of 2PCF, 
\begin{equation}
  \delta \widehat{W}_{\rm 2d}^{\alpha, \rm corr}(s, \mu)    = \delta \widehat{W}_{\rm 2d}^{\alpha}(z_1, z_2, s, \mu) - \delta \widehat{W}^{\alpha}_{\rm 2d, {\rm sys}}(z_1, z_2, s, \mu) \label{eq:correlation_w_2d}.
\end{equation}

\subsection{PCA data compression }\label{sec:method_PCA}

In principle, utilizing the full shape of the correlation function can maximize the information regarding constraints on cosmology. For instance, the findings of \citep{Park:2019mvn} and \citep{Dong_2023} demonstrate that the $(s,\mu)$-dependent 2-dimensional 2PCF significantly improves compared to the 1-dimensional $\mu$-dependent 2PCF. Consequently, we will also employ 2-dimensional MCFs to constrain cosmology. However, a computational challenge arises in the estimation of the covariance. The accurate estimation of covariance for a large number of statistical quantities depends on a substantial number of mock samples, requiring extensive simulation and computational time.

When the number of mock catalogs is limited, a common strategy is to adopt coarser binning. This involves averaging the data over a reduced number of bins, which helps stabilize the covariance matrix and enables more efficient and robust $\chi^2$ evaluations. In this study, we compress 116 finer $\mu$-bins in the range $0 < \mu < 0.97$ into 18 coarser bins. However, a key drawback of this straightforward bin-averaging compression scheme is the loss of information, which can weaken the constraining power of the analysis.

To address this issue, we propose a principal component analysis (PCA)-based compression scheme that operates directly on the unbinned data, supported by \texttt{scikit-learn}\citep{scikit-learn}. This approach enables efficient $\chi^2$ computation while preserving as much information as possible from the full shape of MCFs. PCA reduces the dimensionality of a dataset by transforming correlated variables into an orthogonal set of uncorrelated modes, retaining those with the largest eigenvalues. This minimizes correlations while preserving key information, improving the feasibility and robustness of the analysis. We assume that only the first $N_c$ principal components contain relevant cosmological information, with $N_c$ determined through simulations based on explained variance ratio $r_c$. See details in Sec.\ref{sec:result_PCA}.

It's a compression algorithm similar to PCA called "massive optimized parameter estimation and data compression" (MOPED, \citep{MOPED}). The algorithm is lossless in theory because it keeps the Fisher information matrices identical. We try to introduce it into our analysis. Although there are some problems and limitations, it still shows great potential. We intend to make a detailed analysis in the future work, while only do preliminary tests and comparisons with PCA in this work. See details in Appendix \ref{sec:appendix_MOPED}.

In general, the PCA compression procedure is performed through the following steps. The $N_p \times N_p$ empirical covariance matrix $\boldsymbol{C}$ (Note that it's not the same meaning as $\boldsymbol{C}$ in MCMC in this work) of a vector $\boldsymbol{p}$ is estimated using the following procedure:
\begin{equation}\label{eq:cov}
\boldsymbol{C} \equiv \frac{1}{N_{\rm mock}-1} \sum_{i=1}^{N_{\rm mock}} \left(\boldsymbol{p}_i - \boldsymbol{\bar{p}}\right) \left(\boldsymbol{p}_i-\boldsymbol{\bar{p}}\right)^T\,,
\end{equation}
where the summation is taken over all $N_{\rm mock}$ mock samples. Here, $\boldsymbol{p}_i$, a vector of length $N_p$, represents the statistical quantities for the $i$-th mock sample. Note that the mean of $\boldsymbol{p}$ over all mocks is denoted by $\boldsymbol{\bar{p}}$. The next step involves diagonalizing the covariance matrix $\boldsymbol{C}$ through a standard singular value decomposition (SVD), achieved as follows.
\begin{equation}
\boldsymbol{C} = \boldsymbol{U} \boldsymbol{\Lambda} \boldsymbol{U}^T\,.
\end{equation}
Here $\boldsymbol{U}$ is an orthogonal matrix such that $\boldsymbol{U}\boldsymbol{U}^T = \boldsymbol{I}$, where its columns serve as the eigenvectors. Meanwhile, the matrix $\boldsymbol{\Lambda}$ takes the form of a diagonal matrix, with its elements denoted as $\Lambda_{jk} = \delta_{jk}\lambda_j$, where $\lambda_j$ represents the eigenvalues arranged in descending order.

When retaining only the first $N_c$ dominant eigenmodes, the projection is performed using the first $N_c$ columns of the matrix $\boldsymbol{U}$, denoted as $\boldsymbol{U}_c$ with dimensions $N_p \times N_c$ ($N_c \ll N_p$). The transformation of the data vectors $\boldsymbol{p}_i$ and $\boldsymbol{a}_i$ is given by: \begin{equation}\label{eq:transform} \boldsymbol{a}_i = \boldsymbol{U}_c^T(\boldsymbol{p}_i-\boldsymbol{\bar{p}})\,,\quad \boldsymbol{p}'_i = \boldsymbol{U}_c\boldsymbol{a}_i\,, \end{equation} 
where the data is compressed from $N_p$ to $N_c$ dimensions while preserving most of the relevant information.

In addition, elements in $\boldsymbol{a}$ are uncorrelated in the sense that 
\begin{equation}
\langle\boldsymbol{a}_i\boldsymbol{a}_i^T\rangle= \boldsymbol{U}_c^T \boldsymbol{C} \boldsymbol{U}_c =\boldsymbol{\Lambda}_c\,,
\end{equation}
where $\boldsymbol{\Lambda}_c$ is a diagonal matrix that keeps only the first $N_c$ diagonal components of $\boldsymbol{\Lambda}$. By applying PCA dimensionality reduction to the data, we have maintained a significant amount of cosmological information with a small number of vector lengths. The choice of $N_c$ is determined through a convergence test, which will be detailed in Sect.~\ref{sec:results}.



To train the PCA for covariance compression, as described in Eq.~\ref{eq:cov}, we derive the measured quantities for each statistic from a set of mock simulations. Each mock is generated using specific values of $\Omega_m$ and $w$, with the full set of mocks sampling a wide range of these parameters: $\Omega_m \in [0.2, 0.5]$ and $w \in [-1.5, 0.5]$. In total, we use $100 \times 100$ mock datasets, uniformly spaced across the $\Omega_m$–$w$ parameter space.

Using the relevant parameters for each grid point, the measured statistics are mapped directly from fiducial cosmologies to other cosmologies with values for these parameters according to the mapping scheme of Eq.~\ref{eq:coordinate}. Furthermore, we assume that the redshift evolution of anisotropy from RSD is insensitive to the choice of $\Omega_m$--$w$ within the specified range of interest. We further discuss it in Sec.\ref{sec:con}. Consequently, the mapping can span the entire parameter space, allowing for the efficient generation of the required statistics in large numbers.



\section{Analysis and result}\label{sec:results}
In this section, we examine how assuming an incorrect cosmology affects the measured statistics, quantifying the sensitivity of deviations over the range $\alpha \in [-0.3, 1]$. We present constraints on $\Omega_m$ and $w$ from different statistics, showing that combining MCFs at multiple $\alpha$s yields the strongest constraints. Additionally, we demonstrate that the proposed PCA compression scheme significantly enhances constraint power by more effectively extracting information than the conventional coarse binning approach.

Furthermore, we assess the impact of redshift errors, a realistic systematic in future stage-IV slitless spectroscopic surveys. We find that redshift uncertainties significantly weaken the constraints from the 2PCF, but only slightly affect those from the combined MCFs. Notably, when using PCA compression, the cosmological constraints remain largely unaffected by this systematic. 

At the end of this section, we specifically added robustness testing to highlight the limitations of our method at present. Specifically, when processing gull 2-dimensional 2PCF and MCFs, due to the overly simplistic PCA compression strategy currently adopted, the results are highly dependent on the accuracy of systematic estimation. This will be the part that we will focus on improving in our future work.

\subsection{Comparasion between 2PCF and MCFs for the coarse binning scheme}\label{sec:results_coarse}
\begin{figure*}[htbp]
    \centering
    \includegraphics[width=0.95 \textwidth]{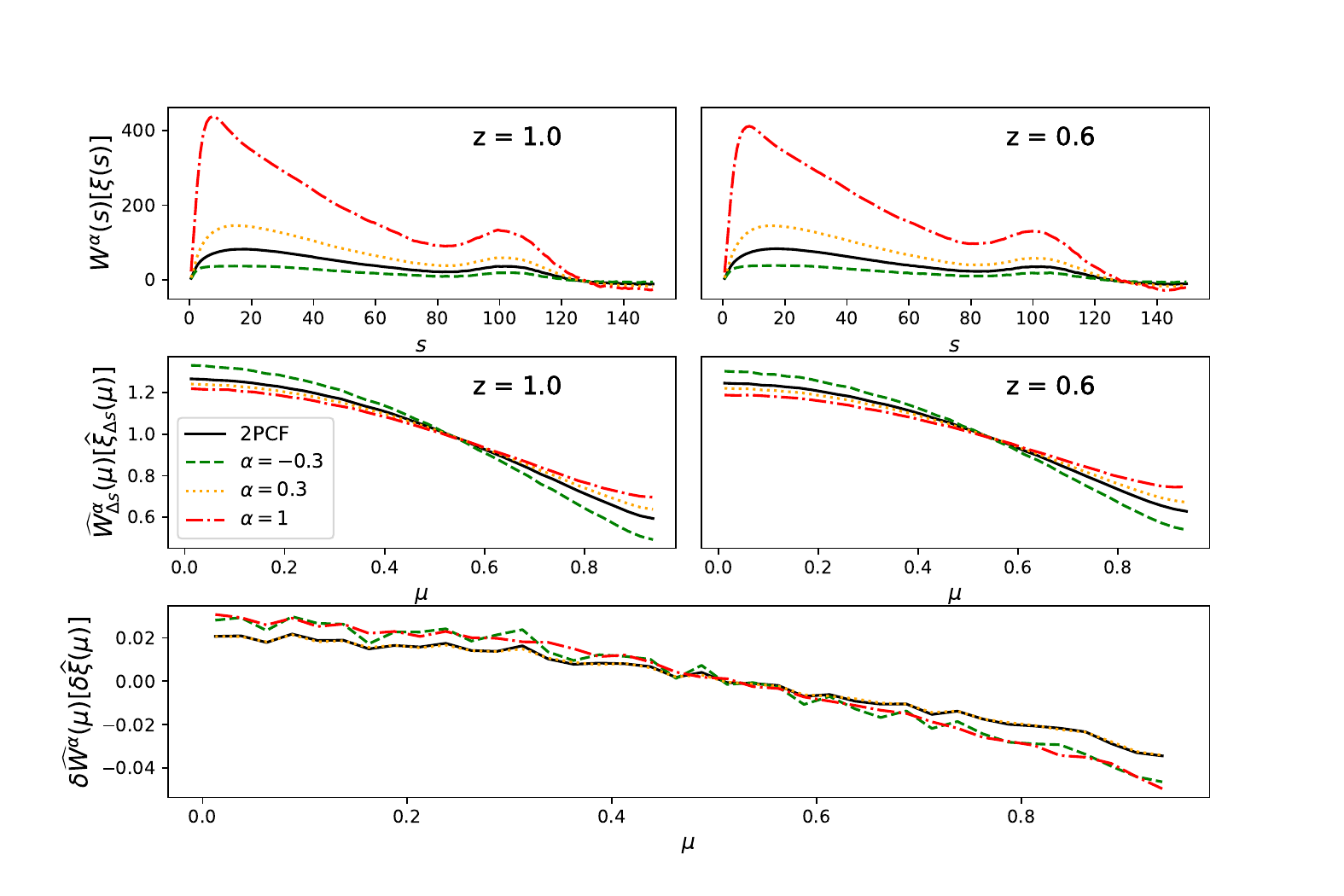}
    \caption{Comparison between the standard 2PCF and four MCFs with $\alpha = -0.3$, $0.3$, and $1$ at $z = 1$ and $z = 0.6069$, showing the radial (upper) and angular (middle) dependence. Besides, the lower panel shows the evolution of angular dependence of the standard 2PCF and these three MCFs. $W_{\Delta \mu}^{\alpha}(s) \equiv \int_{0}^{\mu_{\mathrm{cut}}} W^{\alpha}(s, \mu) \mathrm{d}\mu$, where $\mu_{\mathrm{max}} = 0.97$, while $\hat{W}_{\Delta s}^{\alpha}(\mu)$ is defined as Eq.\ref{eq:w_normalization}. The correspondence between 2PCF and MCFs remains consistent with the line styles and colors on each panel.}
    \label{fig:Xis_MCFs}
\end{figure*}

Compared to the standard 2PCF, MCFs reflect the halo density environment, with different $\alpha$s being more sensitive to either dense or sparse regions. The 1-dimensional 2PCF and MCFs, integrated over $\mu$ and $s$ at $z = 1.0$, are shown in Fig.~\ref{fig:Xis_MCFs}, where $W_{\Delta \mu}^{\alpha}(s) \equiv \int_{0}^{\mu_{\mathrm{max}}} W^{\alpha}(s, \mu) \mathrm{d}\mu$ with $\mu_{\mathrm{max}} = 0.97$, while $\hat{W}_{\Delta s}^{\alpha}(\mu)$ is defined as Eq.\ref{eq:w_normalization}. To suppress the FOG effect, we exclude the region $\mu > 0.97$. 

As shown in Fig.\ref{fig:Xis_MCFs}, the 1-dimensional anisotropic MCFs ($\widehat{W}_{\Delta s}^{\alpha}(\mu)$) are sensitive to $\alpha$. For $\alpha < 0$, radial inhomogeneity decreases and angular anisotropy increases, while for $\alpha > 0$, the opposite occurs. By combining different values of $\alpha$ in the MCFs, we can extract more information than from each individual value of $\alpha$ alone, thereby providing stronger constraints in the tomographic AP test. Note that the difference between $\delta \widehat{\xi}_{\Delta s}(\mu)$ and $\delta \widehat{W}^{0.3}_{\Delta s}(\mu)$ is extremely small, implying that there is a high correlation between these two cases. Therefore, we do not combine them in the analysis.

\begin{figure*}[htbp]
    \centering
    \includegraphics[width=0.9 \textwidth]{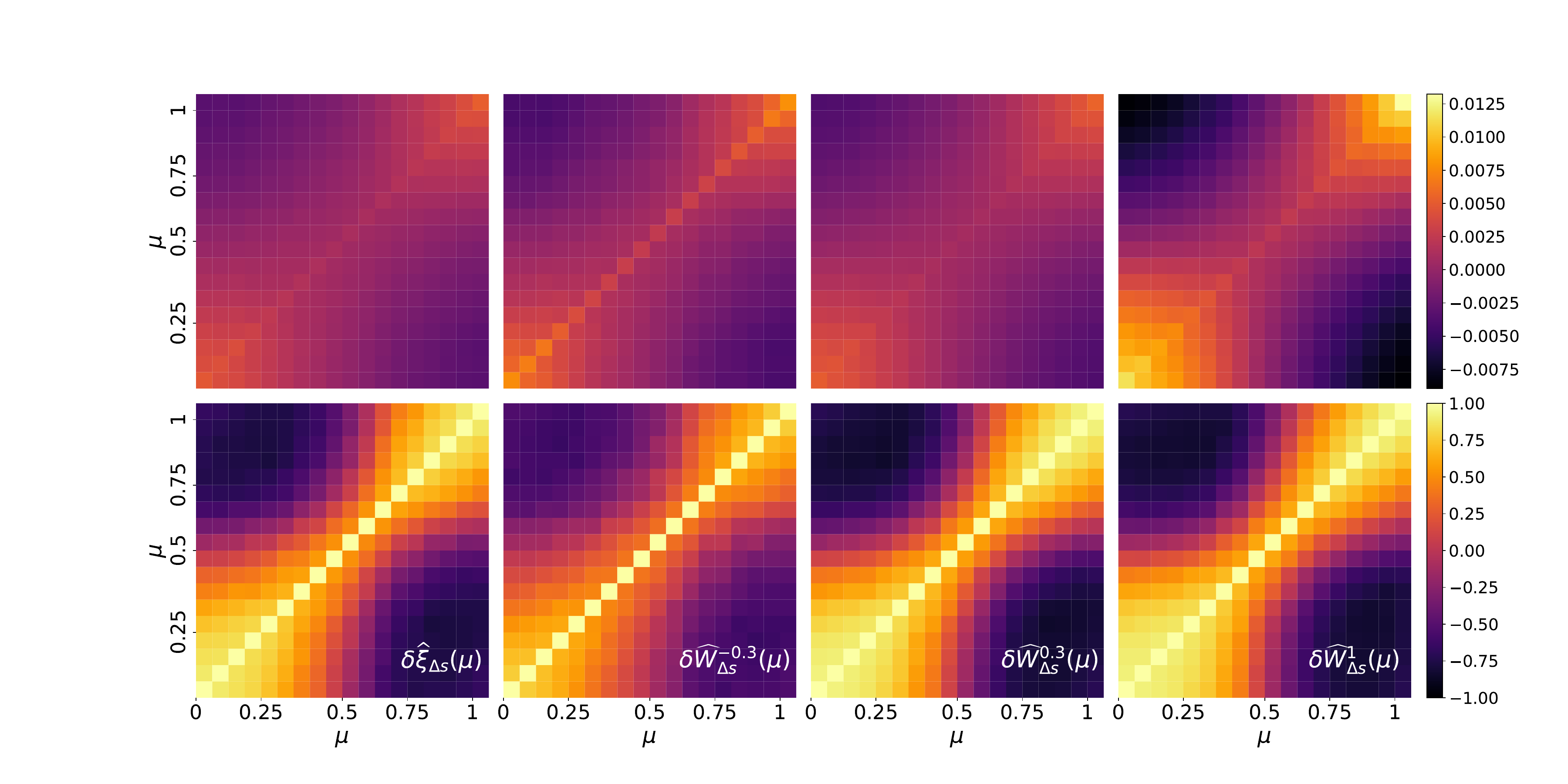}
    \caption{Covariance matrices (top) and correlation coefficients (bottom) for the 1-dimensional standard 2PCF and MCFs with $\alpha = -0.3$, $0.3$, and $1$, respectively.}
    \label{fig:cov_corr_2PCF_MCFs}
\end{figure*}

\begin{figure}[htbp]
    \centering
    \includegraphics[width= 0.45\textwidth]{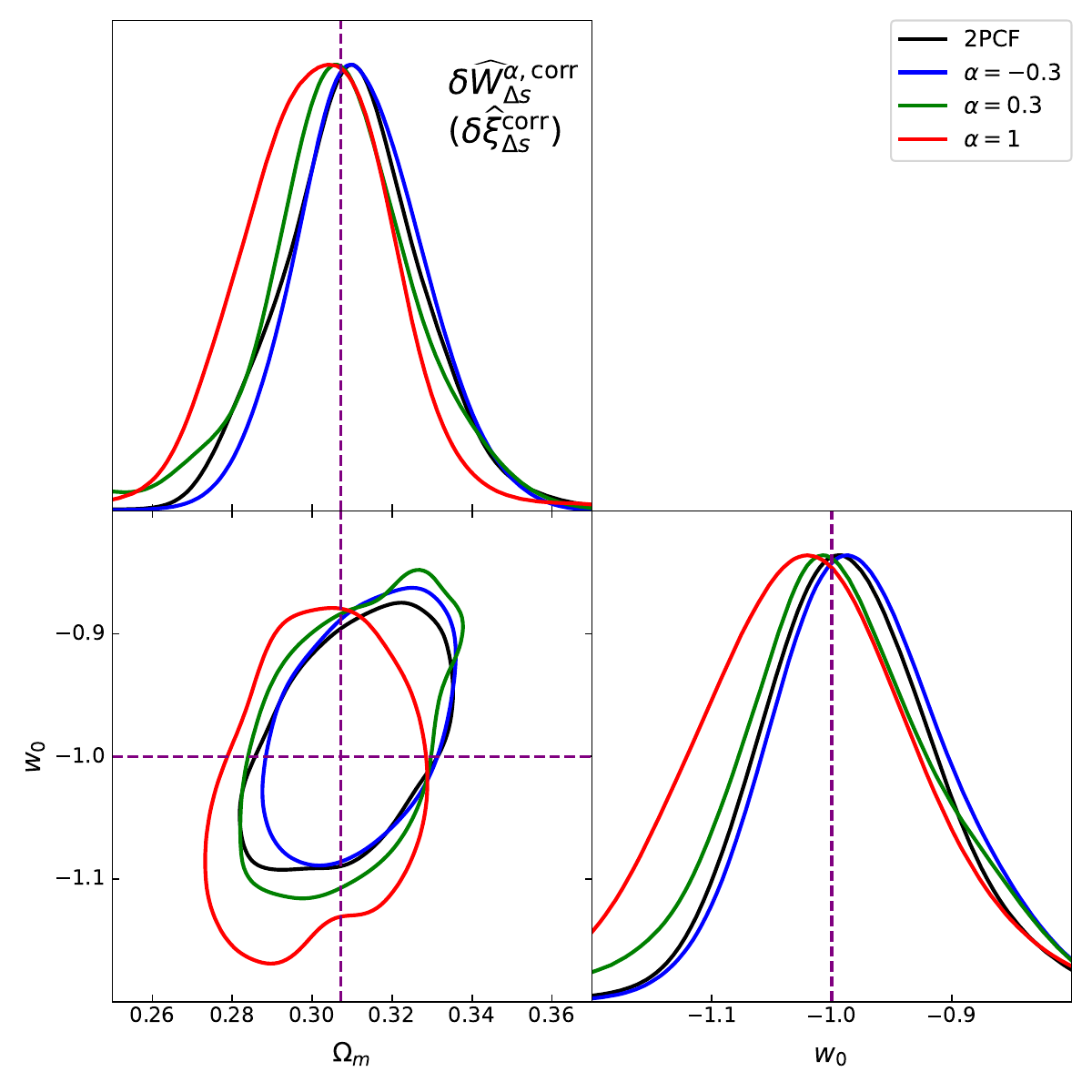}
    \caption{Comparison of the posterior distributions of the cosmological parameters $\Omega_m$ and $w$ at the 68\% confidence level, derived from the standard 2PCF and MCFs with $\alpha = -0.3$, $0.3$, and $1$, based on MCMC analysis, where $\delta \widehat{W}_{\Delta s}^{\alpha, \rm corr}$ is defined as Eq.\ref{eq:correction_W}, while $\delta \widehat{\xi}_{\Delta s}^{\rm corr}$ is defined as Eq.\ref{eq:correlation_xi}. The fiducial values $\Omega_m = 0.3071$ and $w = -1.0$ are indicated by the dashed lines.}
    \label{fig:MCMC_results}
\end{figure}

We now present the constraint results on $\Omega_m$ and $w$ obtained from combined MCFs with a coarse binning scheme, and in the following subsection, demonstrate that the constraining power can be enhanced by adopting a finer binning scheme combined with PCA compression.  Note that in this work we only consider 1-dimensional 2PCF and combined MCFs with the coarse binning method because the it is extremely inefficient for applying to 2-dimensional 2PCF and combined MCFs, which include massive correlations between different bins and bring great difficulties to the estimation of covariance.

The covariance matrices and correlation coefficients of the 1-dimensional standard 2PCF and MCFs are shown in Fig.~\ref{fig:cov_corr_2PCF_MCFs}. As evident, there exist strong correlations between different bins. This motivates the use of PCA compression, especially when adopting a finer binning scheme ($n = 116$), where the number of bins is approximately 6 times greater than in the coarse binning case ($n=17$).

Based on the covariance matrices, we compute the likelihoods for the standard 2PCF and for the three MCFs corresponding to different $\alpha$s: $-0.3, 0.3$ and $1$. These likelihoods are then used to derive posterior distributions of the cosmological parameters $\Omega_m$ and $w$ using MCMC sampling. The resulting parameter constraints are shown in Fig.~\ref{fig:MCMC_results}.

As shown, the overall constraining power of the 2PCF and MCFs is comparable, and none of the individual $\alpha$ values alone significantly reduce the parameter space. However, from the joint $\Omega_m$–$w$ posterior distributions, we observe that the case with $\alpha = 1$ yields notably different constraints compared to the others. Therefore, as pointed out by~\cite{Yang:2020ysv}, we will combine the 2PCF with MCFs at the four $\alpha$ values in the following analysis and demonstrate that this leads to improved constraints.

\begin{figure}[!htbp]
    \centering
    \includegraphics[width= 0.45\textwidth]{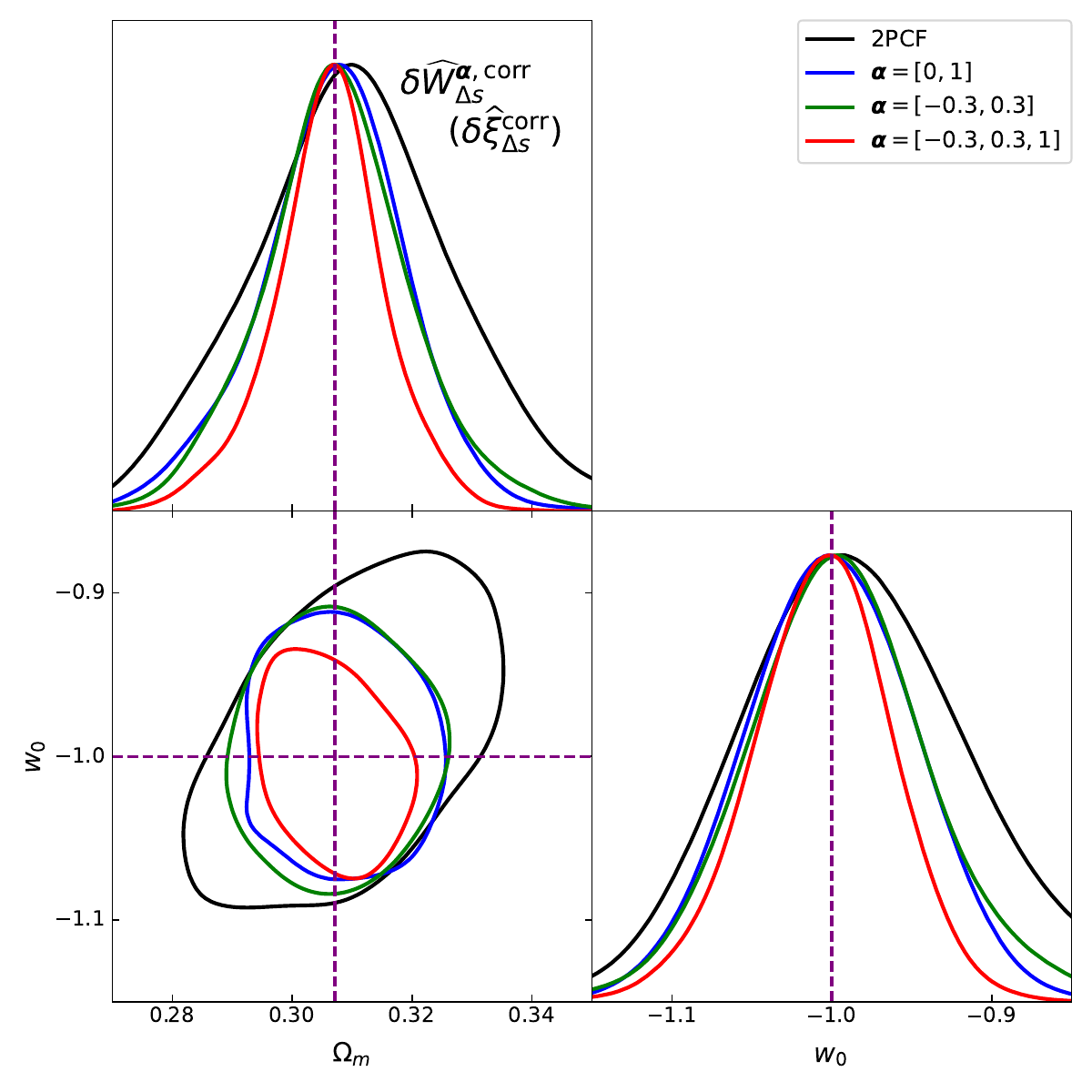}
    \caption{Same as in Fig.~\ref{fig:MCMC_results}, but for combined MCFs with two or three $\alpha$s for illustration: $\boldsymbol{\alpha} = [0, 1]$, $\boldsymbol{\alpha} = [-0.3, -0.1]$, and $\boldsymbol{\alpha} = [-0.3, -0.1, 0.3]$. Clearly, the marginalized uncertainties are reduced relative to the 2PCF baseline.}
    \label{fig:combination_MCFs}
\end{figure}

Three examples of combined MCFs ($\delta W^{\boldsymbol{\alpha}}_{\Delta s}$) are considered for illustration: specifically, $\boldsymbol{\alpha} = [0, 1]$, $\boldsymbol{\alpha} = [-0.3, -0.1]$, and $\boldsymbol{\alpha} = [-0.3, -0.1, 0.3]$. The resulting constraints from the combined MCFs are shown in Fig.~\ref{fig:combination_MCFs}. As shown, the statistical uncertainties in the marginalized distributions of $\Omega_m$  ($w$) are reduced by 34.9 (34.8)\%, 31.1 (27.6)\%, and 48.9 (45.2)\% for the three respective MCF combinations, relative to the 2PCF case.

\subsection{Improved constraints from MCFs using PCA compression}\label{sec:result_PCA}

In the following, we will demonstrate the improvements in the constraints obtained from the PCA compression for the fine binning scheme, as compared to the simpler coarse binning scheme. The PCA models are trained with the samples including AP signals instead of those including cosmic variance and structural fluctuations, trying to avoid the PCA models recognizing noise as signal and leading to overfitting.  Our focus will be on comparing the constraints derived from the 1-dimensional angular 2PCF and MCFs.

\subsubsection{Constraints from 1D angular 2PCF and MCFs}

\begin{figure*}[htbp]
    \centering
    \includegraphics[width=0.9 \textwidth]{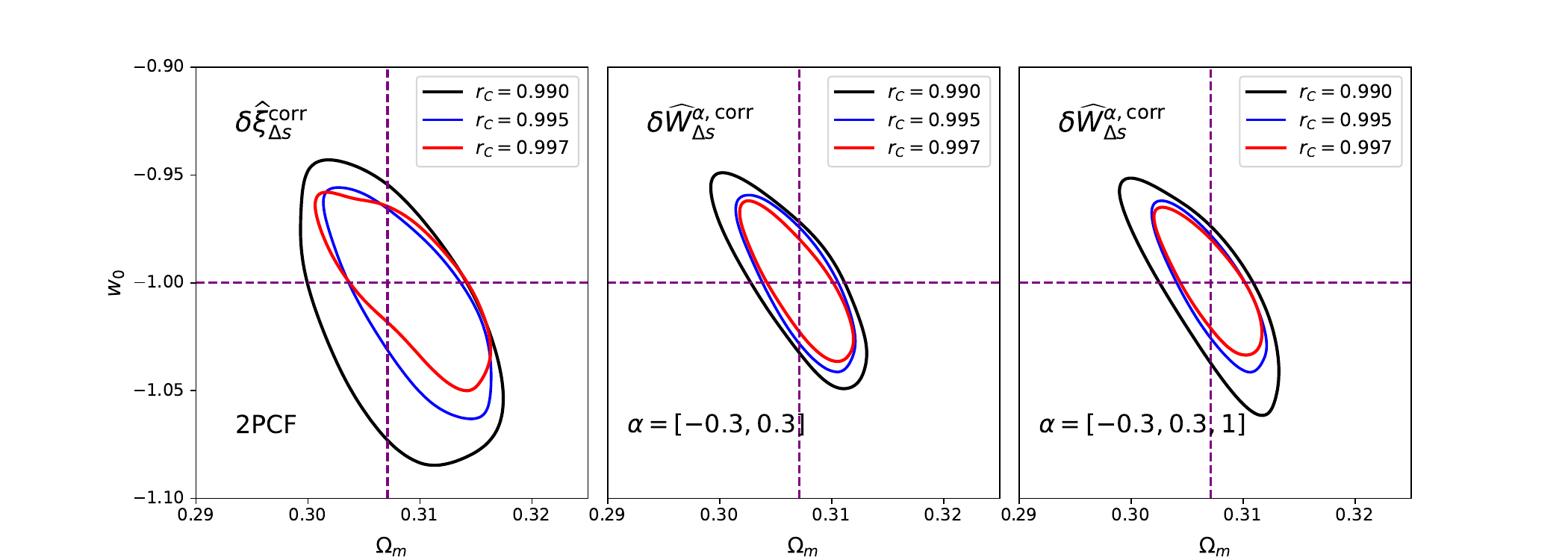}
    \caption{Convergence test based on the constraints on the $\Omega_m$--$w$ parameter space. From left to right: results for the 2PCF, and the combined MCFs with $\boldsymbol{\alpha} = [-0.3, 0.3]$ and $\boldsymbol{\alpha} = [-0.3, 0.3, 1]$, respectively. In each panel, we show results for $r_c = 0.99$, $0.995$, and $0.997$. Increasing $r_c$ from 0.99 to 0.995 noticeably affects the constraints, while further increasing $r_c$ yields negligible improvement.} 
    \label{fig:PCA_convergence}
\end{figure*}

Due to the limited number of mock realizations, the covariance matrices cannot be accurately estimated, particularly for eigenvectors associated with small or near-zero eigenvalues. This leads to numerical instability when inverting the covariance matrix during likelihood evaluations. The coarse binning method mentioned earlier can achieve a reduction in covariance, however, leading to significant information loss. To address this issue, we apply PCA to reduce the dimensionality of the covariance matrix while retaining as much information as possible that is helpful for cosmological constraints. Specifically, we keep only the dominant eigenvectors corresponding to large eigenvalues where the signal-to-noise ratio is high, and discard those dominated by noise, which are poorly estimated by the limited mocks. This PCA-based compression significantly reduces the dimensionality of the data, improving numerical stability and the robustness of the likelihood analysis.

In practice, we select an appropriate cutoff $ N_c$ and keep only the first to the $N_c$-th eigenvectors, assuming the eigenvalues are sorted in descending order. The choice of $N_c$ is guided by the explained variance ratio, ensuring that the selected components capture a sufficient amount of information. Specifically, we require the cumulative variance to satisfy $\sum_{i=1}^{N_c} \lambda_i^2/\sum_i \lambda_i^2  \geq r_c$, where we adopt $r_c = 0.995$, ensuring that at least 99.5\% of the total variance is preserved by the selected $N_c$ eigenvectors.

To assess whether the chosen explained variance ratio $r_c$ is appropriate for our analysis, we perform a convergence test. By varying the explained variance threshold and comparing the resulting parameter constraints, we can evaluate the stability of the results. If the constraints remain consistent as the explained variance ratio increases, this indicates convergence and supports the suitability of the selected threshold.
 
In Fig.~\ref{fig:PCA_convergence}, we present the constraints on the $\Omega_m$--$w$ parameter space. From left to right, we show the results for the 2PCF case, and the combined MCFs with $\boldsymbol{\alpha} = [-0.3, 0.3]$ and $\boldsymbol{\alpha} = [-0.3, 0.3, 1]$, respectively. In each panel, we choose $r_c = 0.99$, 0.995, and 0.997 respectively. The responding $N_c$ are different, and as an example, they are $N_c = 11$, 17 and 24 for the $\boldsymbol{\alpha} = [-0.3, 0.3, 1]$ case.

As shown, increasing $r_c$ from 0.99 to 0.995 leads to noticeable changes in the parameter constraints. However, further increasing $r_c$ from 0.995 to 0.997 results in only slight improvements. The blue and red contours are nearly overlapping, indicating that the constraints have effectively converged. Quantitatively, the differences in the 1-dimensional marginalized uncertainties for each parameter in each case are less than 10\%. Thus, we adopt $r_c = 0.995$ for this analysis.

\begin{figure}[!htbp]
    \centering
    \includegraphics[width=0.45 \textwidth]{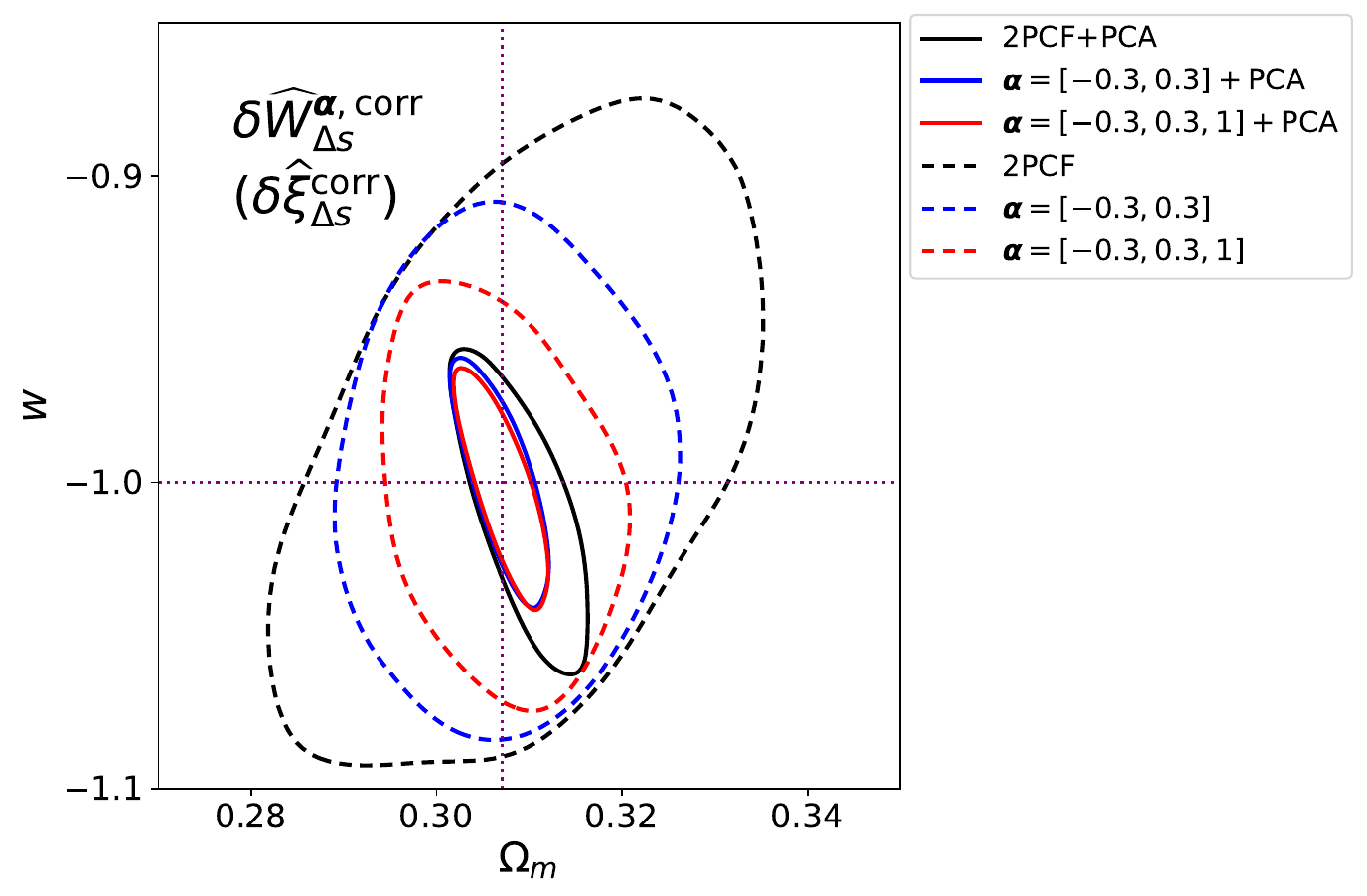}
    \caption{Joint distributions for $\Omega_m$ and $w$ dervied from the the statistical quantity $\delta \widehat{\xi}_{\Delta s}^{\rm corr}$ and $\delta \widehat{W}_{\Delta s}^{\alpha, \rm corr}$ for different binning scheme. Solid lines show results from fine binning with PCA compression and $r_c = 0.995$, while dotted lines represent coarse binning. The fine binning+PCA method significantly improves constraints by effectively extracting the information from a large number of bins, compared to the coarse binning scheme. Combined MCFs with $\alpha = [-0.3, 0.3, 1]$ yield the tightest constraints, followed by $\alpha = [-0.3, 0.3]$, while the 2PCF alone provides the weakest.}
    \label{fig:MCMC_PCA_comparison}
\end{figure}

In Fig.~\ref{fig:MCMC_PCA_comparison}, we observe that, in general, the fine binning scheme combined with PCA compression (solid lines) yields significantly tighter constraints compared to the coarse binning scheme (dotted lines). This improvement arises from the effective extraction of information from a large number of fine bins via PCA.

Furthermore, we find that the 1-dimensional 2PCF alone provides weaker constraining power than the combined MCFs with multiple $\alpha$s. The most stringent constraints on $\Omega_m$ and $w$ are obtained by combining $\boldsymbol{\alpha} = [-0.3, 0.3, 1]$ with PCA compression, while using only $\boldsymbol{\alpha} = [-0.3, 0.3]$ results in slightly weaker constraints. Specifically, we find $\Omega_m = 0.3071 \pm 0.0034$ ($0.3070 \pm 0.0035$) and $w = -1.0020 \pm 0.0261$ ($-1.0001 \pm 0.0275$) for the three (two)  $\alpha$s case, respectively, and $\Omega_m = 0.3092 \pm 0.0050$, $w = -1.0093 \pm 0.0358$ for the 2PCF case.

\subsubsection{Constraints from 2-dimensional 2PCF and MCFs using PCA}

For the full  2-dimensional 2PCF and MCFs, there are $34 \times 116 = 3944$ bins (34 bins in $s$ and 116 bins in $\mu$) for each of them. Due to the high dimensionality of these fine bins and the limited number of mock realizations, it becomes challenging to accurately estimate the covariance matrix. Therefore, we apply PCA to compress the data and extract the signal-dominated information.

For convenience and as a preliminary testing, we apply PCA compression directly to all bins without any other compression methods in this work. PCA is applied to the 2-dimensional 2PCF ($\delta \widehat{\xi}_{\rm 2d}^{\rm corr}(z_1, z_2, s, \mu)$, Eq.\ref{eq:correlation_xi_2d}) and two different schemes of combined MCFs ($\delta \widehat{W}_{\rm 2d}^{\alpha, \rm corr}(z_1, z_2, s, \mu)$, Eq.\ref{eq:correlation_w_2d}): $\boldsymbol{\alpha} = [-0.3, 0.3]$ and $\boldsymbol{\alpha} = [-0.3, 0.3, 1]$. We find that the convergence test is satisfied with $N_c \sim 30$ components in both cases, using a threshold of $r_c = 0.95$, which compresses the data vector from $7888 (\boldsymbol{\alpha} = [-0.3, 0.3])$ or $11832 (\boldsymbol{\alpha} = [-0.3, 0.3, 1])$ to about $30$ dimensions. Increasing the threshold to $r_c = 0.956$ yields essentially no change in the resulting constraints. The mean values of $\Omega_m$ and $w$ shift by approximately 0.01\%, with the associated $1\sigma$ error bars varying by approximately 5\%. We do not fix $r_c = 0.995$ because of $N_c \sim 80$ with it in the case of 2-dimensional 2PCF or MCFs, which leads to extremely difficult covariance estimation. Besides, it's unnecessary because of the convergence test of $r_c = 0.95$. 

\begin{figure}[!htbp]
    \centering
    \includegraphics[width=0.45 \textwidth]{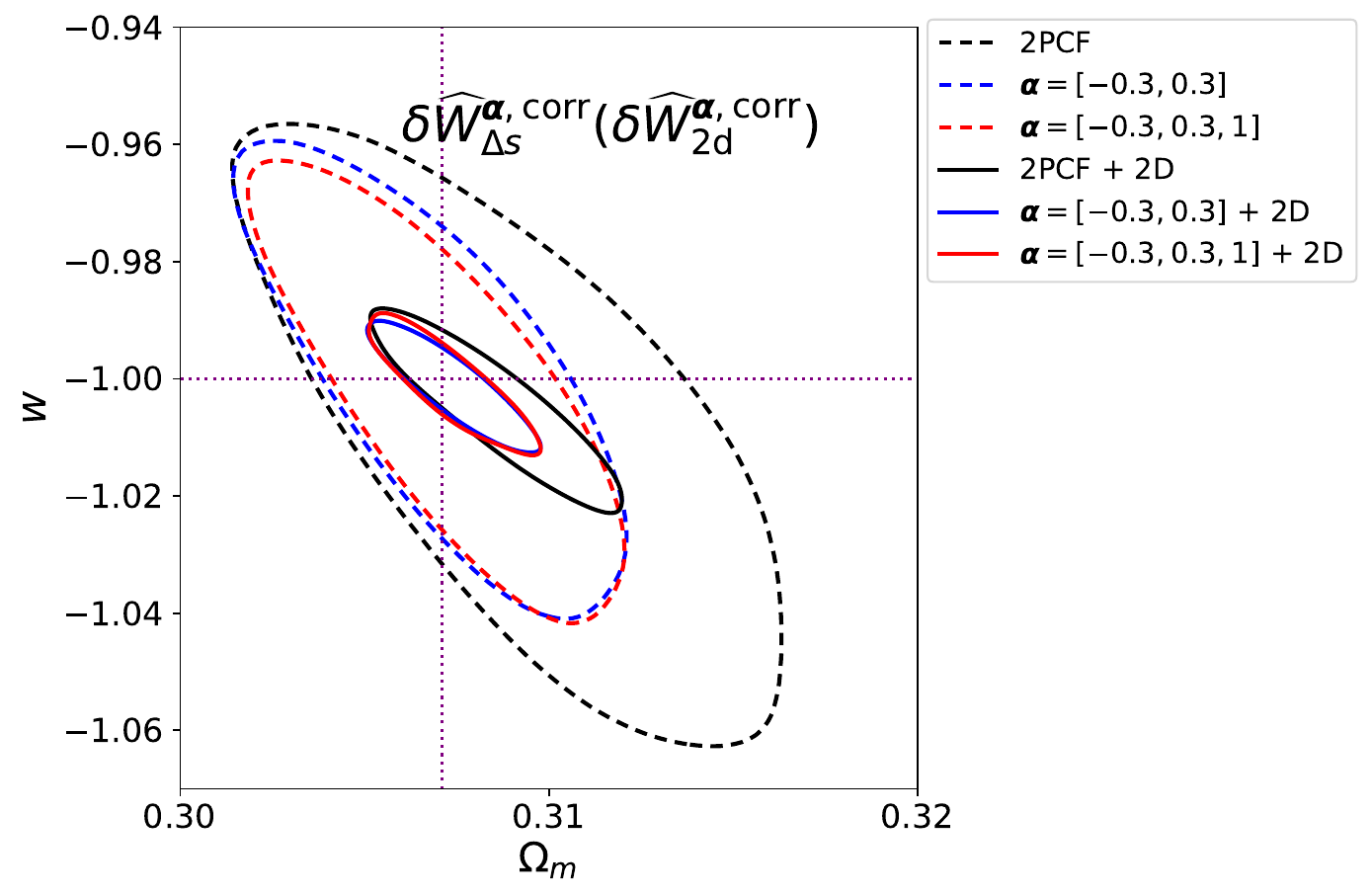}
    \caption{ Comparison of  constraints from 1D and 2D 2PCFs and MCFs, all derived using the fine binning scheme with PCA. For a fair comparison, we adopt $r_c = 0.995$ for the 1D case and $r_c = 0.95$ for the 2D case to ensure convergence. The full-shape information in both 2PCFs and MCFs significantly improves the constraints compared to the 1D case. Both combined MCFs provide substantially tighter constraints than the 2PCF alone.}
    \label{fig:MCMC_contour_2D_1D}
\end{figure}

To further highlight the improvement brought by the full shape of the MCFs,  we present  the constraint results derived from the 1D and 2D 2PCFs and MCFs in Fig.~\ref{fig:MCMC_contour_2D_1D}, each computed using the fine binning scheme with PCA. For a fair comparison, we adopt $r_c = 0.995$ for the 1D case and $r_c = 0.950$ for the 2D case, ensuring that the results are well converged. As shown, the full shape information in both the 2PCFs and MCFs significantly enhances the constraining power compared to the 1D statistics. Moreover, the combined MCFs with $\boldsymbol{\alpha} = [-0.3, 0.3, 1]$ yield the most stringent constraints. It should be noted that the contours from $\boldsymbol{\alpha} = [-0.3, 0.3]$ closely overlap with those from $\boldsymbol{\alpha} = [-0.3, 0.3, 1]$, indicating that adding more $\alpha$s yields negligible improvement. More importantly, either of the combined MCF configurations provides significantly tighter constraints than the 2PCF alone.

\begin{figure*}[!htbp]
    \centering
    \includegraphics[width=0.95 \textwidth]{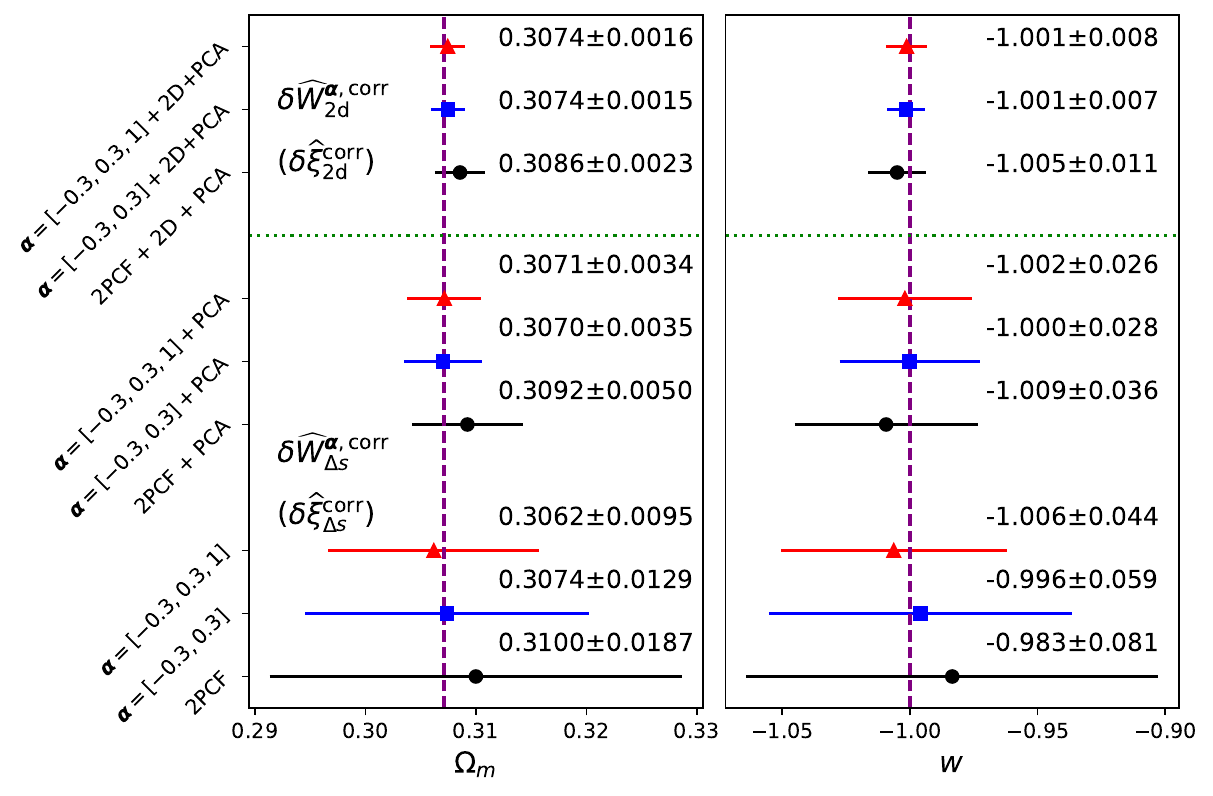}
    \caption{Comparison of the constraining power on $\Omega_m$ and $w$ across different cases, as labeled on the $y$-axis. PCA compression is applied with $r_c = 0.995$ for 1D and $r_c = 0.95$ for 2D statistics to ensure convergence. The constraints of 2D and 1D statistics are seperated by the green dotted line.  From top to bottom, the results correspond to: i) 2D statistics with fine binning and PCA; ii) 1D statistics with fine binning and PCA; and iii) 1D statistics with coarse binning without PCA. In each case, the constraints from the combined MCFs with $\boldsymbol{\alpha} = [-0.3, 0.3]$ and $[-0.3, 0.3, 1]$ are compared to those from the 2PCF.}
    \label{fig:MCMC_errorbar}
\end{figure*}

Fig.~\ref{fig:MCMC_errorbar} presents a comparison of the constraining power from various cases, as labeled on the $y$-axis. When PCA compression is applied, we adopt $r_c = 0.995$ and $0.95$ for the 1D and 2D cases, respectively, to ensure the convergence of the constraints. From top to bottom, we illustrate the results from: i) full-shape statistics with fine binning and PCA compression; ii) 1D statistics with fine binning and PCA compression; and iii) 1D statistics with coarse binning without PCA compression. In each case, we also compare the results from the combined MCFs with $\alpha = [-0.3, 0.3, 1]$ and $[-0.3, 0.3]$ to those from the 2PCF.

From the results, it can be seen that the full-shape statistics with PCA compression provide significantly stronger constraints than the other methods. The combined MCFs further enhance the constraining power, yielding even tighter constraints than the 2PCF. The 1D statistics with PCA compression yield relatively weaker constraints than 2D statistics, while the 1D statistics with coarse binning perform the worst among all cases. Quantitatively, the statistical uncertainty of the strongest constraint is reduced by approximately 85\% compared to the weakest one. 

However, it has to be noted that it's just a preliminary test of the full-shape statistics because we simply train PCA models with all $s$ and $\mu$ bins without additional processing, which is likely to result in high sensitivity to specific systematics. A robustness check on it is shown in Sec.\ref{sec:robustness_SYS} and a significant bias of constraint is shown if we have the bias of systematics estimation. It will be necessary to conduct more in-depth research in the future to eliminate or weaken this high sensitivity to systemic effects.

\subsection{Impact of redshift errors}

 In future stage-IV slitless spectroscopic surveys, redshift errors will be non-negligible, reducing resolution along the line of sight and potentially compromising the tomographic AP method~\citep{Xiao_2022}. These errors can mimic a false FOG effect, blurring both the AP and RSD signals. Redshift uncertainties distort cosmological measurements by biasing the inferred comoving distances, quantified as $\sigma_r = c \sigma_z / H(z)$. As a result, clustering information on scales $\lesssim \sigma_r$ becomes inaccessible. For instance, the upcoming CSST optical survey expects a redshift error of $\sigma_z \sim 0.002$, an order of magnitude larger than current spectroscopic surveys, corresponding to a few Mpc uncertainty in comoving distance.

To evaluate the impact of redshift errors on cosmological parameter estimation, we model them as
\begin{equation} \label{eq:RE} 
\sigma_z = \sigma(1+z)^\gamma\,,
\end{equation}
where $\sigma$ and $\gamma$ are free parameters. We consider three representative cases: $(\sigma, \gamma) = [0.002, 1.0]$, $[0.005, 1.0]$, and $[0.002, 1.5]$. The first reflects typical redshift errors expected from the CSST survey, while the latter two--one with larger amplitude and the other with a steeper redshift dependence--are used to test the robustness of the tomographic AP method.

\begin{figure*}[!htpb]
    \centering
    \includegraphics[width=0.95 \textwidth]{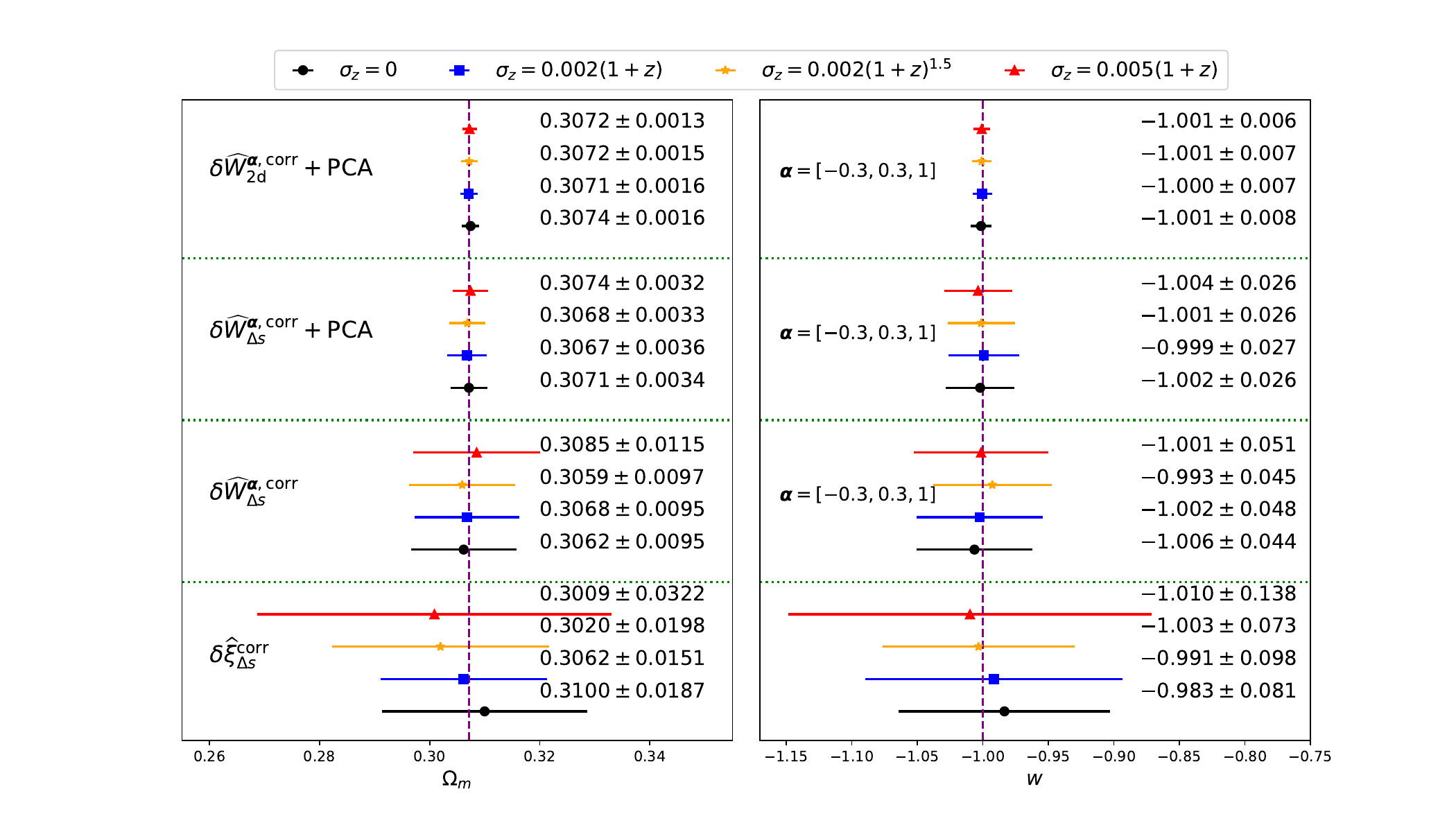}
    \caption{Summary of the parameter estimates for $\Omega_m$ (left) and $w$ (right) using various methods to validate robustness against redshift errors. Across all redshift-error models with ``2D MCFs + PCA'' method, deviations from the redshift-error-free case (black) are negligibly small, with shifts in mean values and $1\sigma$ uncertainties limited to about 0.1\% and 10\%. The ``1D MCFs + PCA'' method shows similar stability, further supporting the reliability of our method. In contrast, directly applying coarse binning to the 2PCF and 1D MCFs leads to large fluctuations in estimates. This highlights the effectiveness of PCA in mitigating redshift-error contamination by preserving high signal-to-noise eigenmodes and suppressing noise-dominated ones.}
    \label{fig:MCMC_errorbar_RE}
\end{figure*}

The results are presented in Fig.~\ref{fig:MCMC_errorbar_RE}. We adopt  a new strategy to decide cutoff thresholds. Specifically, we fix the explained variance ratio of the last eigenvector with it in the corresponding redshift-error-free cases. We adopt this strategy because we believe that PCA plays a significant role in denoising in this section.  For the MCFs, we fix the combined $\alpha$ values to $\boldsymbol{\alpha} = [-0.3, 0.3, 1]$, as this configuration was previously shown to provide the strongest constraints in the case of coarse binning without PCA compression while in the case of fine binning with PCA compression, it was shown to provide the similar constraints to $\boldsymbol{\alpha} = [-0.3,0.3]$.

As shown, the parameter estimates for both $\Omega_m$ and $w$ using the ``2D MCFs + PCA'' scheme remain robust against redshift errors. The differences compared to the redshift-error-free case (shown in black) are negligibly small across all redshift-error models, with mean values and $1\sigma$ uncertainties changing by only about 0.1\% and 10\%. Similarly, for the "1D MCFs + PCA" case, the changes remain at the level of about 0.2\% and 20\%, further demonstrating the robustness of our method.

In contrast, directly applying the coarse binning scheme to the 2PCF and 1D MCF statistics results in large fluctuations in both the estimated mean values and associated uncertainties. This comparison highlights the effectiveness of the PCA technique in mitigating redshift-error-induced contamination, as it preserves the most informative high signal-to-noise modes while suppressing the impact of noise-dominated eigenmodes.

\subsection{Robustness check about the systematics estimation}\label{sec:robustness_SYS}

A key point of tomographic AP test is the accuracy of systematic estimation. 
We will not investigate this complex issue in depth here, since the systematic uncertainties depend critically on each survey's data analysis methodology. Besides, \citep{LI18} found that the change in the cosmological constraints is not significant($\lesssim 0.2 \sigma$ for a SDSS-like survey) even  ignoring the systematic effect and do not correct it in the joint-analysis with 2PCF. 

In any case, since we adopt a new methodology based on MCFs and PCA, it is necessary to conduct a preliminary assessment of systematic uncertainties, to gain at least a qualitative understanding.
In this section, we examine two scenarios where the systematic error estimation suffer from some mistakes due to: (1) the use of incorrectly biased tracers, 2) a slightly mis-specified redshift error model.

\subsubsection{The effect of tracer bias}
To investigate the effect of tracer bias on the systematic estimation, we change the halos and subhalos mass thresholds to: 
\begin{equation}
    m_H \in 
    \begin{cases}
    [3.0\times10^{12}, 2.6 \times 10^{13}] M_{\odot}/h, &  z = 1.0 \\ 
    [3.5 \times 10^{12}, 2.2 \times 10^{13}] M_{\odot}/h, & z = 0.6069 \\ 
    \end{cases}\label{eq:new_mass_threshold}
\end{equation}

\noindent fixing number density as $10^{-3} (\mathrm{Mpc}/h)^{-3}$.

We expect this change affects the redshift evolution of RSD (i.e. corresponding to the items including the suffix "sys" in Eq. \ref{eq:correlation_xi}, Eq.\ref{eq:correction_W}, Eq.\ref{eq:correlation_xi_2d} and Eq.\ref{eq:correlation_w_2d}), and thus has some impact on the derived cosmological constraints. In detail, we re-calculate the redshift evolution of RSD of combined MCFs with $\boldsymbol{\alpha} = [-0.3,0.3,1]$, in the case of coarse binning and PCA compression, and run the MCMC process again. 

\begin{figure*}[!htpb]
    \centering
    \includegraphics[width=0.95 \textwidth]{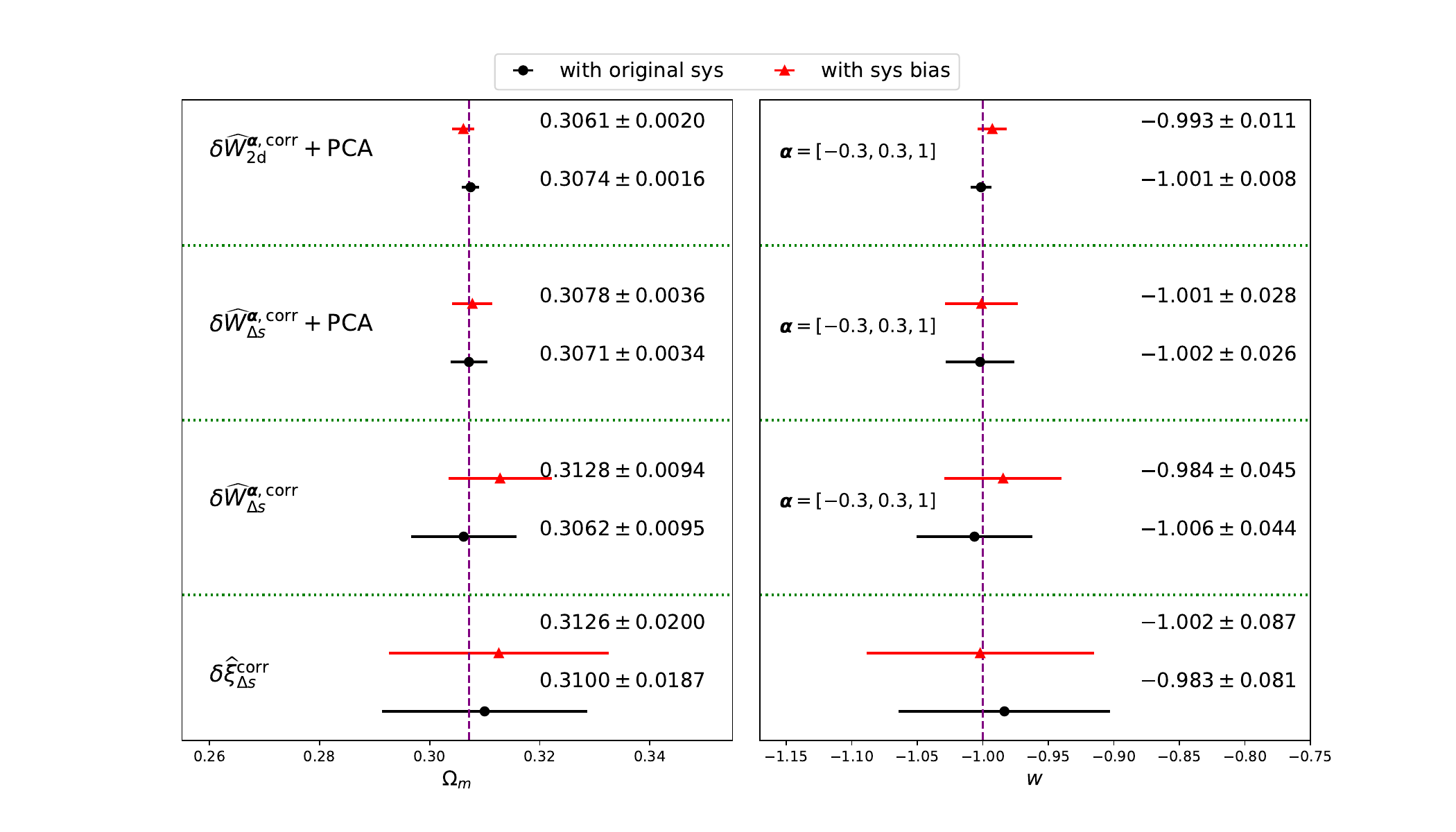}
    \caption{The effect on $\Omega_m-w$ constraints from different halos and subhalos mass cut thresholds. The errorbars with black lines and dots denote the cosmological contraints($1-\sigma$) with original systematics while the errorbars with red lines and triangles denote those with systematic bias from the new halos and subhalos cut thresholds (Eq.\ref{eq:new_mass_threshold}). For convenience, we separate four parts with green dotted lines, corresponding to the four statistical vectors: 1) $\mu-$dependence 2PCF with coarse binning approach, 2) combined $\mu-$dependence MCFs with coarse binning approach, 3) combined $\mu-$dependence MCFs with PCA compression, and 4) combined $(s,\mu)-$dependence MCFs with PCA compression. All combined MCFs are combined with $\boldsymbol{\alpha} = [-0.3, 0.3, 1]$. The constraints of $\Omega_m$ and $ w $ are on the left and right panels, respectively. The purple dash lines denote the fiducial values of $\Omega_m = 0.3071$ and $w = -1.0$.}
    \label{fig:MCMC_massSYS}
\end{figure*}

The effect on cosmological constraints is shown in Fig.\ref{fig:MCMC_massSYS}. In this case, the effect on $\Omega_m-w$ constraints of different halos and subhalos mass cuts is not significant. Especially, in the case of combined MCFs with PCA compression, the bias between best fit and fiducial values is $\lesssim 0.2 \sigma$. The only exception is that, the bias reaches nearly $1\sigma$ in the case of 2-dimensional combined MCFs ($\delta \widehat{W}^{\boldsymbol{\alpha}, \mathrm{corr}}_{\mathrm{2d}}$). In this case, the cosmological constraints are very tight, so a small effect could alter the final results with high statistical confidence.

\subsubsection{Mis-specified redshift errors}

In this section, we examine the effect on constraints of mis-specified redshift errors, with the form of $\sigma_z = 0.002(1+z)^{1.01}$ while the fiducial redshift errors are $\sigma_z = 0.002(1+z)$. 

\begin{figure*}[!htpb]
    \centering
    \includegraphics[width=0.9 \textwidth]{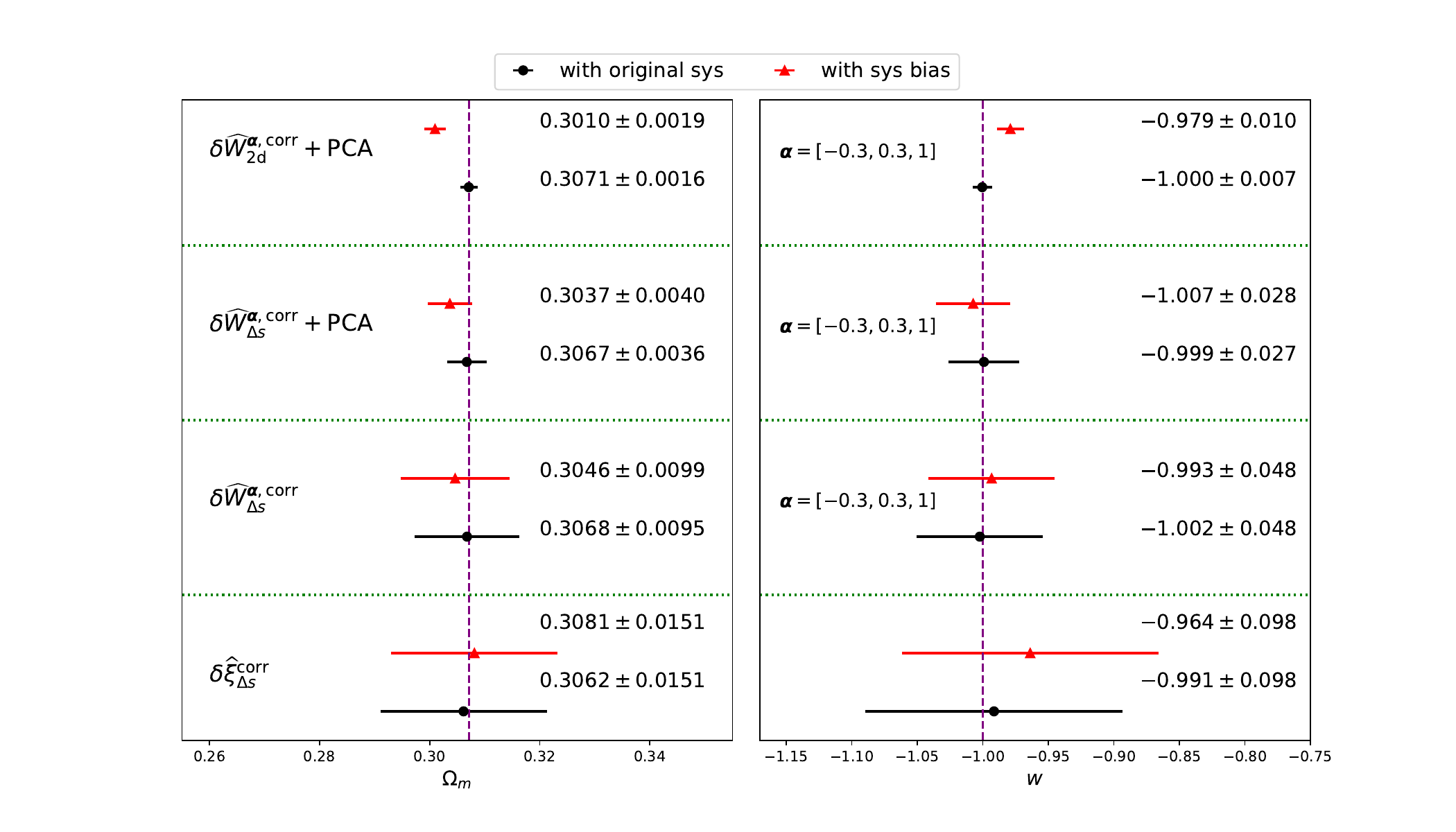}
    \caption{The effect on $\Omega_m-w$ constraints from mis-specified redshift errors. The lines and markers represent the same meaning as Fig.\ref{fig:MCMC_massSYS}.}
    \label{fig:MCMC_reSYS}
\end{figure*}

The effect on cosmological constraints is shown in Fig.\ref{fig:MCMC_reSYS}. Still, except the case of 2-dimensional combined MCFs with PCA compression, the impact is limited (almost $\lesssim 0.3 \sigma$). The case of 2-dimensional combined MCFs shows significant impact from mis-specified redshift errors, the bias of $\Omega_m$ and $w$ reaches $2\sigma$ and $3\sigma$, which implies that the current simple PCA compression strategy is not robust against mis-specified redshift errors. 

In summary, we find that cosmological constraints derived using the PCA components of 2-dimensional MCFs are sensitive to tracer bias inaccuracies and redshift error mis-specification. In the future, we will do more works to enhance the robustness of the methodological and address these limitations.

\section{Summary and discussion}\label{sec:con}

The Alcock-Paczynski (AP) test serves as a powerful tool for probing the expansion history of the Universe. This method can be applied to any randomly oriented cosmological structures whose ensemble average exhibits spherical symmetry. By statistically analyzing distortions in the large-scale galaxy distribution, the AP effect provides a robust means of constraining key cosmological parameters. A notable advancement, the tomographic AP method, effectively mitigates contamination from redshift-space distortions (RSD) by leveraging the redshift evolution of large-scale structure (LSS) anisotropy. Since this anisotropy is sensitive to the AP effect but not to RSD, the technique enables a clean separation between the two types of distortions. 

In this work, we introduce the marked correlation functions (MCFs) into the analysis procedure of the tomographic AP test to enhance its ability to constrain cosmological parameters. Three statistical measures--$\delta \widehat{{\xi}}_{{\Delta s}}^{{\rm corr}}(\mu)$, $\delta \widehat{{W}}_{{\Delta s}}^{{\boldsymbol{\alpha}, \rm corr}}(\mu)$ and $\delta \widehat{{W}}_{{\rm 2d}}^{{\boldsymbol{\alpha}, \rm corr}}(s, \mu)$-- are used to quantify anisotropic clustering. We find that combining different weights ($\alpha = [-0.3, 0.3, 1]$) in MCFs yields the tightest constraints on the cosmological parameters $\Omega_m$ and $w$. 

To maximize information extraction from anisotropic clustering measurements, we implement a Principal Component Analysis (PCA) compression scheme. This method efficiently projects high-dimensional statistical data onto a reduced set of eigenmodes while preserving the majority of cosmological information. Our analysis reveals that conventional coarse binning strategies—employed to maintain tractable covariance matrices—substantially degrade the precision of parameter constraints. In contrast, PCA compression demonstrates superior performance, particularly for the 2-dimensional marked correlation functions $\delta \widehat{{W}}_{{\rm 2d}}^{{\boldsymbol{\alpha}, \rm corr}}(s, \mu)$, where the large parameter space makes dimensionality reduction especially impactful.

By using the full shape of MCFs, i.e. $\delta \widehat{{W}}_{{\rm 2d}}^{{\boldsymbol{\alpha}, \rm corr}}(s, \mu)$ obtained by combining multiple $\alpha$s, we significantly reduce the statistical uncertainties in $\Omega_m$ and $w$ by factors of 30.4\% and 27.3\% compared to the standard 2PCF, respectively. The most stringent constraints from the combined-weight MCFs are $\Omega_m = 0.3074 \pm 0.0016$ and $w = -1.001 \pm 0.008$ for $\alpha = [-0.3, 0.3, 1]$. However, we must emphasize the following fact: the PCA compression strategy used for the full shape of MCFs is too simple, which leads to a strong sensitivity on systematics estimation, meaning that the constraint power in this area can only serve as a potential manifestation. We will conduct more in-depth research on reducing its sensitivity in the future.

Furthermore, to test the robustness of our analysis, we incorporate redshift errors expected in future stage-IV slitless spectroscopic surveys. Using three redshift error models, we find that the full-shape MCFs combined with PCA yield highly robust constraints -- even under substantial redshift uncertainty such as $\sigma_z = 0.005(1+z)$ --comparable to the redshift-error-free case. 

In particular, it should be noted that the systematic estimation is unbiased in this work because we utilize the same mock in the whole analysis. It's an important problem in the analysis of surveys. A simple robustness check on the problem of our method is shown in Sec. \ref{sec:robustness_SYS} and it replies that, except for the full-shape of 2PCF or MCFs, the impact is limited. We plan to do the further check in the futural work.

Our study demonstrates how to effectively extract large-scale structure (LSS) information using the tomographic Alcock-Paczyński (AP) method with MCFs. These techniques can be adapted to other surveys and datasets, potentially helping to constrain additional cosmological parameters. The framework also shows strong potential for future emulator-based analyses with the China Space Station Telescope (CSST). We provide a simple extension of this work to CSST in Appendix \ref{sec:appendix_extension} based on the relation of covariance and galaxy number density\citep{LI19}. Besides, according to the number density and survey volume from ~\citep{Miao_2022}, the constraints could be around 30\% tighter than those in this work for a CSST-like survey, without considering other errors. We plan to do a formal preparation in the future, when the mock with accurate systematic estimation of CSST is ready. 

In the future, we plan to expand our analysis by considering the methodology developed in ~\cite{XiaoYuan_2022}, which suggests that the inclusion of the density gradient as a weight may further improve the sensitivity to cosmological parameters. In addition, we do not consider the validation of redshift evolution of RSD in different cosmologies. Some tests have been done, showing that cosmological parameters do have an impact on the RSD evolution but the impact is small \citep{Park:2019mvn}. With the improvement of the accuracy of the stage-IV surveys, it is indeed necessary to consider the validation. We are going to do the validation in the future, when the mock with different $w \mathrm{CDM}$ cosmologies that meet our requirement is ready. We also emphasize that, although this work focuses on MCFs, the tomographic AP method is applicable to almost any kind of statistics, as long as the statistics is capable of probing anisotropic clustering. This makes it an extremely promising tool for exploring nonlinear scales. We plan to continue investigating this approach in future studies.

\section*{Acknowledgements}

This work is supported by the Ministry of Science and Technology of China (2020SKA0110401, 2020SKA0110402, 2020SKA0110100), the National Key Research and Development Program of China (2018YFA0404504, 2018YFA0404601, 2020YFC2201600), the National Natural Science Foundation of China (12373005, 11890691, 12205388, 12220101003, 12473097), the China Manned Space Project with numbers CMS-CSST-2021 (A02, A03, B01), Guangdong Basic and Applied Basic Research Foundation (2024A1515012309) and the Fundamental Research Funds for the Central Universities, Sun Yat-sen University(No. 24qnpy122). This work was performed using the Tianhe-2 supercomputer and the Kunlun cluster in the School of Physics and Astronomy at Sun Yat-Sen University. We also wish to acknowledge the Beijing Super Cloud Center (BSCC) and Beijing Beilong Super Cloud Computing Co., Ltd (http://www.blsc.cn/) for providing HPC resources that have significantly contributed to the research results presented in this paper.


\section*{Data availability}
The $\mathtt{BigMD}$ simulation used in this paper is available via the CosmoSim database (https://www.cosmosim.org/).


\bibliographystyle{mod-apsrev4-2}
\bibliography{cites_new2.bib}

@ARTICLE{DESI:2024mwx,
       author = {{Adame}, A.~G. and {Aguilar}, J. and {Ahlen}, S. and {Alam}, S. and {Alexander}, D.~M. and {Alvarez}, M. and {Alves}, O. and {Anand}, A. and {Andrade}, U. and {Armengaud}, E. and {Avila}, S. and {Aviles}, A. and {Awan}, H. and {Bahr-Kalus}, B. and {Bailey}, S. and {Baltay}, C. and {Bault}, A. and {Behera}, J. and {BenZvi}, S. and {Bera}, A. and {Beutler}, F. and {Bianchi}, D. and {Blake}, C. and {Blum}, R. and {Brieden}, S. and {Brodzeller}, A. and {Brooks}, D. and {Buckley-Geer}, E. and {Burtin}, E. and {Calderon}, R. and {Canning}, R. and {Carnero Rosell}, A. and {Cereskaite}, R. and {Cervantes-Cota}, J.~L. and {Chabanier}, S. and {Chaussidon}, E. and {Chaves-Montero}, J. and {Chen}, S. and {Chen}, X. and {Claybaugh}, T. and {Cole}, S. and {Cuceu}, A. and {Davis}, T.~M. and {Dawson}, K. and {de la Macorra}, A. and {de Mattia}, A. and {Deiosso}, N. and {Dey}, A. and {Dey}, B. and {Ding}, Z. and {Doel}, P. and {Edelstein}, J. and {Eftekharzadeh}, S. and {Eisenstein}, D.~J. and {Elliott}, A. and {Fagrelius}, P. and {Fanning}, K. and {Ferraro}, S. and {Ereza}, J. and {Findlay}, N. and {Flaugher}, B. and {Font-Ribera}, A. and {Forero-S{\'a}nchez}, D. and {Forero-Romero}, J.~E. and {Frenk}, C.~S. and {Garcia-Quintero}, C. and {Gazta{\~n}aga}, E. and {Gil-Mar{\'\i}n}, H. and {Gontcho a Gontcho}, S. and {Gonzalez-Morales}, A.~X. and {Gonzalez-Perez}, V. and {Gordon}, C. and {Green}, D. and {Gruen}, D. and {Gsponer}, R. and {Gutierrez}, G. and {Guy}, J. and {Hadzhiyska}, B. and {Hahn}, C. and {Hanif}, M.~M.~S. and {Herrera-Alcantar}, H.~K. and {Honscheid}, K. and {Howlett}, C. and {Huterer}, D. and {Ir{\v{s}}i{\v{c}}}, V. and {Ishak}, M. and {Juneau}, S. and {Kara{\c{c}}ayl{\i}}, N.~G. and {Kehoe}, R. and {Kent}, S. and {Kirkby}, D. and {Kremin}, A. and {Krolewski}, A. and {Lai}, Y. and {Lan}, T. -W. and {Landriau}, M. and {Lang}, D. and {Lasker}, J. and {Le Goff}, J.~M. and {Le Guillou}, L. and {Leauthaud}, A. and {Levi}, M.~E. and {Li}, T.~S. and {Linder}, E. and {Lodha}, K. and {Magneville}, C. and {Manera}, M. and {Margala}, D. and {Martini}, P. and {Maus}, M. and {McDonald}, P. and {Medina-Varela}, L. and {Meisner}, A. and {Mena-Fern{\'a}ndez}, J. and {Miquel}, R. and {Moon}, J. and {Moore}, S. and {Moustakas}, J. and {Mueller}, E. and {Mu{\~n}oz-Guti{\'e}rrez}, A. and {Myers}, A.~D. and {Nadathur}, S. and {Napolitano}, L. and {Neveux}, R. and {Newman}, J.~A. and {Nguyen}, N.~M. and {Nie}, J. and {Niz}, G. and {Noriega}, H.~E. and {Padmanabhan}, N. and {Paillas}, E. and {Palanque-Delabrouille}, N. and {Pan}, J. and {Penmetsa}, S. and {Percival}, W.~J. and {Pieri}, M.~M. and {Pinon}, M. and {Poppett}, C. and {Porredon}, A. and {Prada}, F. and {P{\'e}rez-Fern{\'a}ndez}, A. and {P{\'e}rez-R{\`a}fols}, I. and {Rabinowitz}, D. and {Raichoor}, A. and {Ram{\'\i}rez-P{\'e}rez}, C. and {Ramirez-Solano}, S. and {Rashkovetskyi}, M. and {Ravoux}, C. and {Rezaie}, M. and {Rich}, J. and {Rocher}, A. and {Rockosi}, C. and {Roe}, N.~A. and {Rosado-Marin}, A. and {Ross}, A.~J. and {Rossi}, G. and {Ruggeri}, R. and {Ruhlmann-Kleider}, V. and {Samushia}, L. and {Sanchez}, E. and {Saulder}, C. and {Schlafly}, E.~F. and {Schlegel}, D. and {Schubnell}, M. and {Seo}, H. and {Shafieloo}, A. and {Sharples}, R. and {Silber}, J. and {Slosar}, A. and {Smith}, A. and {Sprayberry}, D. and {Tan}, T. and {Tarl{\'e}}, G. and {Taylor}, P. and {Trusov}, S. and {Ure{\~n}a-L{\'o}pez}, L.~A. and {Vaisakh}, R. and {Valcin}, D. and {Valdes}, F. and {Vargas-Maga{\~n}a}, M. and {Verde}, L. and {Walther}, M. and {Wang}, B. and {Wang}, M.~S. and {Weaver}, B.~A. and {Weaverdyck}, N. and {Wechsler}, R.~H. and {Weinberg}, D.~H. and {White}, M. and {Yu}, J. and {Yu}, Y. and {Yuan}, S. and {Y{\`e}che}, C. and {Zaborowski}, E.~A. and {Zarrouk}, P. and {Zhang}, H. and {Zhao}, C. and {Zhao}, R. and {Zhou}, R. and {Zhuang}, T.},
        title = "{DESI 2024 VI: cosmological constraints from the measurements of baryon acoustic oscillations}",
      journal = {\jcap},
         year = 2025,
        month = feb,
       volume = {2025},
       number = {2},
          eid = {021},
        pages = {021},
          doi = {10.1088/1475-7516/2025/02/021},
archivePrefix = {arXiv},
       eprint = {2404.03002},
 primaryClass = {astro-ph.CO},
       adsurl = {https://ui.adsabs.harvard.edu/abs/2025JCAP...02..021A}
}

@ARTICLE{DESI:2025zgx,
       author = {{DESI Collaboration} and {Abdul-Karim}, M. and {Aguilar}, J. and {Ahlen}, S. and {Alam}, S. and {Allen}, L. and {Allende Prieto}, C. and {Alves}, O. and {Anand}, A. and {Andrade}, U. and {Armengaud}, E. and {Aviles}, A. and {Bailey}, S. and {Baltay}, C. and {Bansal}, P. and {Bault}, A. and {Behera}, J. and {BenZvi}, S. and {Bianchi}, D. and {Blake}, C. and {Brieden}, S. and {Brodzeller}, A. and {Brooks}, D. and {Buckley-Geer}, E. and {Burtin}, E. and {Calderon}, R. and {Canning}, R. and {Carnero Rosell}, A. and {Carrilho}, P. and {Casas}, L. and {Castander}, F.~J. and {Cereskaite}, R. and {Charles}, M. and {Chaussidon}, E. and {Chaves-Montero}, J. and {Chebat}, D. and {Chen}, X. and {Claybaugh}, T. and {Cole}, S. and {Cooper}, A.~P. and {Cuceu}, A. and {Dawson}, K.~S. and {de la Macorra}, A. and {de Mattia}, A. and {Deiosso}, N. and {Della Costa}, J. and {Demina}, R. and {Dey}, A. and {Dey}, B. and {Ding}, Z. and {Doel}, P. and {Edelstein}, J. and {Eisenstein}, D.~J. and {Elbers}, W. and {Fagrelius}, P. and {Fanning}, K. and {Fern\textbackslash'andez-Garc\textbackslash'ia}, E. and {Ferraro}, S. and {Font-Ribera}, A. and {Forero-Romero}, J.~E. and {Frenk}, C.~S. and {Garcia-Quintero}, C. and {Garrison}, L.~H.},
        title = "{DESI DR2 Results II: Measurements of Baryon Acoustic Oscillations and Cosmological Constraints}",
      journal = {arXiv e-prints},
         year = 2025,
        month = mar,
          eid = {arXiv:2503.14738},
        pages = {arXiv:2503.14738},
          doi = {10.48550/arXiv.2503.14738},
archivePrefix = {arXiv},
       eprint = {2503.14738},
 primaryClass = {astro-ph.CO},
       adsurl = {https://ui.adsabs.harvard.edu/abs/2025arXiv250314738D}
}

@ARTICLE{DESI:2025fii,
       author = {{Lodha}, K. and {Calderon}, R. and {Matthewson}, W.~L. and {Shafieloo}, A. and {Ishak}, M. and {Pan}, J. and {Garcia-Quintero}, C. and {Huterer}, D. and {Valogiannis}, G. and {Ure{\~n}a-L{\'o}pez}, L.~A. and {Kamble}, N.~V. and {Parkinson}, D. and {Kim}, A.~G. and {Zhao}, G.~B. and {Cervantes-Cota}, J.~L. and {Rohlf}, J. and {Lozano-Rodr{\'\i}guez}, F. and {Rom{\'a}n-Herrera}, J.~O. and {Abdul-Karim}, M. and {Aguilar}, J. and {Ahlen}, S. and {Alves}, O. and {Andrade}, U. and {Armengaud}, E. and {Aviles}, A. and {BenZvi}, S. and {Bianchi}, D. and {Brodzeller}, A. and {Brooks}, D. and {Burtin}, E. and {Canning}, R. and {Carnero Rosell}, A. and {Casas}, L. and {Castander}, F.~J. and {Charles}, M. and {Chaussidon}, E. and {Chaves-Montero}, J. and {Chebat}, D. and {Claybaugh}, T. and {Cole}, S. and {Cuceu}, A. and {Dawson}, K.~S. and {de la Macorra}, A. and {de Mattia}, A. and {Deiosso}, N. and {Demina}, R. and {Dey}, Arjun and {Dey}, Biprateep and {Ding}, Z. and {Doel}, P. and {Eisenstein}, D.~J. and {Elbers}, W. and {Ferraro}, S. and {Font-Ribera}, A. and {Forero-Romero}, J.~E. and {Garrison}, Lehman H. and {Gazta{\~n}aga}, E. and {Gil-Mar{\'\i}n}, H. and {Gontcho}, S. Gontcho A and {Gonzalez-Morales}, A.~X. and {Gutierrez}, G. and {Guy}, J. and {Hahn}, C. and {Herbold}, M. and {Herrera-Alcantar}, H.~K. and {Honscheid}, K. and {Howlett}, C. and {Juneau}, S. and {Kehoe}, R. and {Kirkby}, D. and {Kisner}, T. and {Kremin}, A. and {Lahav}, O. and {Lamman}, C. and {Landriau}, M. and {Le Guillou}, L. and {Leauthaud}, A. and {Levi}, M.~E. and {Li}, Q. and {Magneville}, C. and {Manera}, M. and {Martini}, P. and {Meisner}, A. and {Mena-Fern{\'a}ndez}, J. and {Miquel}, R. and {Moustakas}, J. and {Mu{\~n}oz Santos}, D. and {Mu{\~n}oz-Guti{\'e}rrez}, A. and {Myers}, A.~D. and {Nadathur}, S. and {Niz}, G. and {Noriega}, H.~E. and {Paillas}, E. and {Palanque-Delabrouille}, N. and {Percival}, W.~J. and {Pieri}, Matthew M. and {Poppett}, C. and {Prada}, F. and {P{\'e}rez-Fern{\'a}ndez}, A. and {P{\'e}rez-R{\`a}fols}, I. and {Ram{\'\i}rez-P{\'e}rez}, C. and {Rashkovetskyi}, M. and {Ravoux}, C. and {Ross}, A.~J. and {Rossi}, G. and {Ruhlmann-Kleider}, V. and {Samushia}, L. and {Sanchez}, E. and {Schlegel}, D. and {Schubnell}, M. and {Seo}, H. and {Sinigaglia}, F. and {Sprayberry}, D. and {Tan}, T. and {Tarl{\'e}}, G. and {Taylor}, P. and {Turner}, W. and {Vargas-Maga{\~n}a}, M. and {Walther}, M. and {Weaver}, B.~A. and {Wolfson}, M. and {Y{\`e}che}, C. and {Zarrouk}, P. and {Zhou}, R. and {Zou}, H.},
        title = "{Extended Dark Energy analysis using DESI DR2 BAO measurements}",
      journal = {arXiv e-prints},
         year = 2025,
        month = mar,
          eid = {arXiv:2503.14743},
        pages = {arXiv:2503.14743},
          doi = {10.48550/arXiv.2503.14743},
archivePrefix = {arXiv},
       eprint = {2503.14743},
 primaryClass = {astro-ph.CO},
       adsurl = {https://ui.adsabs.harvard.edu/abs/2025arXiv250314743L}
}

@ARTICLE{EUCLID,
       author = {{Laureijs}, R. and {Amiaux}, J. and {Arduini}, S. and
         {Augu{\`e}res}, J. -L. and {Brinchmann}, J. and {Cole}, R. and
         {Cropper}, M. and {Dabin}, C. and {Duvet}, L. and {Ealet}, A.},
        title = "{Euclid Definition Study Report}",
      journal = {arXiv e-prints},
         year = "2011",
        month = "Oct",
          eid = {arXiv:1110.3193},
        pages = {arXiv:1110.3193},
archivePrefix = {arXiv},
       eprint = {1110.3193},
 primaryClass = {astro-ph.CO},
       adsurl = {https://ui.adsabs.harvard.edu/abs/2011arXiv1110.3193L}
}

@article{Wang_2008,
  title = {Figure of merit for dark energy constraints from current observational data},
  author = {Wang, Yun},
  journal = {Phys. Rev. D},
  volume = {77},
  issue = {12},
  pages = {123525},
  numpages = {7},
  year = {2008},
  month = {Jun},
  publisher = {American Physical Society},
  doi = {10.1103/PhysRevD.77.123525},
  url = {https://link.aps.org/doi/10.1103/PhysRevD.77.123525}
}

@ARTICLE{LI14,
       author = {{Li}, Xiao-Dong and {Park}, Changbom and {Forero-Romero}, J.~E. and
         {Kim}, Juhan},
        title = "{Cosmological Constraints from the Redshift Dependence of the Alcock-Paczynski Test: Galaxy Density Gradient Field}",
      journal = {\apj},
         year = "2014",
        month = "Dec",
       volume = {796},
       number = {2},
          eid = {137},
        pages = {137},
          doi = {10.1088/0004-637X/796/2/137},
archivePrefix = {arXiv},
       eprint = {1412.3564},
 primaryClass = {astro-ph.CO},
       adsurl = {https://ui.adsabs.harvard.edu/abs/2014ApJ...796..137L}
}

@ARTICLE{LI16,
       author = {{Li}, Xiao-Dong and {Park}, Changbom and {Sabiu}, Cristiano G. and
         {Park}, Hyunbae and {Weinberg}, David H. and {Schneider}, Donald P. and
         {Kim}, Juhan and {Hong}, Sungwook E.},
        title = "{Cosmological Constraints from the Redshift Dependence of the Alcock-Paczynski Effect: Application to the SDSS-III BOSS DR12 Galaxies}",
      journal = {\apj},
         year = "2016",
        month = "Dec",
       volume = {832},
       number = {2},
          eid = {103},
        pages = {103},
          doi = {10.3847/0004-637X/832/2/103},
archivePrefix = {arXiv},
       eprint = {1609.05476},
 primaryClass = {astro-ph.CO},
       adsurl = {https://ui.adsabs.harvard.edu/abs/2016ApJ...832..103L}
}

@ARTICLE{LI18,
       author = {{Li}, Xiao-Dong and {Sabiu}, Cristiano G. and {Park}, Changbom and
         {Wang}, Yuting and {Zhao}, Gong-bo and {Park}, Hyunbae and
         {Shafieloo}, Arman and {Kim}, Juhan and {Hong}, Sungwook E.},
        title = "{Cosmological Constraints from the Redshift Dependence of the Alcock-Paczynski Effect: Dynamical Dark Energy}",
      journal = {\apj},
         year = "2018",
        month = "Apr",
       volume = {856},
       number = {2},
          eid = {88},
        pages = {88},
          doi = {10.3847/1538-4357/aab42e},
archivePrefix = {arXiv},
       eprint = {1803.01851},
 primaryClass = {astro-ph.CO},
       adsurl = {https://ui.adsabs.harvard.edu/abs/2018ApJ...856...88L}
}

@ARTICLE{LI19,
       author = {{Li}, Xiao-Dong and {Miao}, Haitao and {Wang}, Xin and {Zhang}, Xue and
         {Fang}, Feng and {Luo}, Xiaolin and {Huang}, Qing-Guo and {Li}, Miao},
        title = "{The Redshift Dependence of the Alcock--Paczynski Effect: Cosmological Constraints from the Current and Next Generation Observations}",
      journal = {\apj},
         year = "2019",
        month = "Apr",
       volume = {875},
       number = {2},
          eid = {92},
        pages = {92},
          doi = {10.3847/1538-4357/ab0f30},
archivePrefix = {arXiv},
       eprint = {1903.04757},
 primaryClass = {astro-ph.CO},
       adsurl = {https://ui.adsabs.harvard.edu/abs/2019ApJ...875...92L}
}

@ARTICLE{Luo_2019,
       author = {{Luo}, Xiaolin and {Wu}, Ziyong and {Li}, Miao and {Li}, Zhigang and {Sabiu}, Cristiano G. and {Li}, Xiao-Dong},
        title = "{Cosmological Constraints from the Redshift Dependence of the Alcock-Paczynski Effect: Fourier Space Analysis}",
      journal = {\apj},
         year = 2019,
        month = dec,
       volume = {887},
       number = {2},
          eid = {125},
        pages = {125},
          doi = {10.3847/1538-4357/ab50b5},
archivePrefix = {arXiv},
       eprint = {1908.10593},
 primaryClass = {astro-ph.CO},
       adsurl = {https://ui.adsabs.harvard.edu/abs/2019ApJ...887..125L}
}

@ARTICLE{Crocce2LPT,
       author = {{Crocce}, Mart{\'\i}n and {Pueblas}, Sebasti{\'a}n and {Scoccimarro}, Rom{\'a}n},
        title = "{Transients from initial conditions in cosmological simulations}",
      journal = {\mnras},
         year = 2006,
        month = nov,
       volume = {373},
       number = {1},
        pages = {369-381},
          doi = {10.1111/j.1365-2966.2006.11040.x},
archivePrefix = {arXiv},
       eprint = {astro-ph/0606505},
 primaryClass = {astro-ph},
       adsurl = {https://ui.adsabs.harvard.edu/abs/2006MNRAS.373..369C}
}

@ARTICLE{BIGMD,
   author = {{Klypin}, A. and {Yepes}, G. and {Gottl{\"o}ber}, S. and {Prada}, F. and 
	{He{\ss}}, S.},
    title = "{MultiDark simulations: the story of dark matter halo concentrations and density profiles}",
  journal = {\mnras},
archivePrefix = "arXiv",
   eprint = {1411.4001},
     year = 2016,
    month = apr,
   volume = 457,
    pages = {4340-4359},
      doi = {10.1093/mnras/stw248},
   adsurl = {https://ui.adsabs.harvard.edu/abs/2016MNRAS.457.4340K}
}

@ARTICLE{ROCKSTAR,
   author = {{Behroozi}, P.~S. and {Wechsler}, R.~H. and {Wu}, H.-Y.},
    title = "{The ROCKSTAR Phase-space Temporal Halo Finder and the Velocity Offsets of Cluster Cores}",
  journal = {\apj},
archivePrefix = "arXiv",
   eprint = {1110.4372},
 primaryClass = "astro-ph.CO",
     year = 2013,
    month = jan,
   volume = 762,
      eid = {109},
    pages = {109},
      doi = {10.1088/0004-637X/762/2/109},
   adsurl = {https://ui.adsabs.harvard.edu/abs/2013ApJ...762..109B}
}

@ARTICLE{AP1979,
   author = {{Alcock}, C. and {Paczynski}, B.},
    title = "{An evolution free test for non-zero cosmological constant}",
  journal = {\nat},
     year = 1979,
    month = oct,
   volume = 281,
    pages = {358},
      doi = {10.1038/281358a0},
   adsurl = {https://ui.adsabs.harvard.edu/abs/1979Natur.281..358A}
}

@ARTICLE{Landy,
   author = {{Landy}, S.~D. and {Szalay}, A.~S.},
    title = "{Bias and variance of angular correlation functions}",
  journal = {\apj},
     year = 1993,
    month = jul,
   volume = 412,
    pages = {64-71},
      doi = {10.1086/172900},
   adsurl = {https://ui.adsabs.harvard.edu/abs/1993ApJ...412...64L}
}

@ARTICLE{Eisenstein:2005su,
   author = {{Eisenstein}, D.~J. and {Zehavi}, I. and {Hogg}, D.~W. and {Scoccimarro}, R. and 
	{Blanton}, M.~R. and {Nichol}, R.~C. and {Scranton}, R. and 
	{Seo}, H.-J. and {Tegmark}, M. and {Zheng}, Z. and {Anderson}, S.~F. and 
	{Annis}, J. and {Bahcall}, N. and {Brinkmann}, J. and {Burles}, S. and 
	{Castander}, F.~J. and {Connolly}, A. and {Csabai}, I. and {Doi}, M. and 
	{Fukugita}, M. and {Frieman}, J.~A. and {Glazebrook}, K. and 
	{Gunn}, J.~E. and {Hendry}, J.~S. and {Hennessy}, G. and {Ivezi{\'c}}, Z. and 
	{Kent}, S. and {Knapp}, G.~R. and {Lin}, H. and {Loh}, Y.-S. and 
	{Lupton}, R.~H. and {Margon}, B. and {McKay}, T.~A. and {Meiksin}, A. and 
	{Munn}, J.~A. and {Pope}, A. and {Richmond}, M.~W. and {Schlegel}, D. and 
	{Schneider}, D.~P. and {Shimasaku}, K. and {Stoughton}, C. and 
	{Strauss}, M.~A. and {SubbaRao}, M. and {Szalay}, A.~S. and 
	{Szapudi}, I. and {Tucker}, D.~L. and {Yanny}, B. and {York}, D.~G.
	},
    title = "{Detection of the Baryon Acoustic Peak in the Large-Scale Correlation Function of SDSS Luminous Red Galaxies}",
  journal = {\apj},
   eprint = {astro-ph/0501171},
     year = 2005,
    month = nov,
   volume = 633,
    pages = {560-574},
      doi = {10.1086/466512},
   adsurl = {https://ui.adsabs.harvard.edu/abs/2005ApJ...633..560E}
}

@ARTICLE{Percival:2007yw,
   author = {{Percival}, W.~J. and {Cole}, S. and {Eisenstein}, D.~J. and 
	{Nichol}, R.~C. and {Peacock}, J.~A. and {Pope}, A.~C. and {Szalay}, A.~S.
	},
    title = "{Measuring the Baryon Acoustic Oscillation scale using the Sloan Digital Sky Survey and 2dF Galaxy Redshift Survey}",
  journal = {\mnras},
archivePrefix = "arXiv",
   eprint = {0705.3323},
     year = 2007,
    month = nov,
   volume = 381,
    pages = {1053-1066},
      doi = {10.1111/j.1365-2966.2007.12268.x},
   adsurl = {https://ui.adsabs.harvard.edu/abs/2007MNRAS.381.1053P}
}

@ARTICLE{Zhang2019,
   author = {{Zhang}, Z. and {Gu}, G. and {Wang}, X. and {Li}, Y.-H. and 
	{Sabiu}, C.~G. and {Park}, H. and {Miao}, H. and {Luo}, X. and 
	{Fang}, F. and {Li}, X.-D.},
    title = "{Nonparametric Dark Energy Reconstruction Using the Tomographic Alcock{--}Paczynski Test}",
  journal = {\apj},
archivePrefix = "arXiv",
   eprint = {1902.09794},
     year = 2019,
    month = jun,
   volume = 878,
      eid = {137},
    pages = {137},
      doi = {10.3847/1538-4357/ab1ea4},
   adsurl = {https://ui.adsabs.harvard.edu/abs/2019ApJ...878..137Z}
}

@ARTICLE{2df:Colless:2003wz,
   author = {{Colless}, M. and {Peterson}, B.~A. and {Jackson}, C. and {Peacock}, J.~A. and 
	{Cole}, S. and {Norberg}, P. and {Baldry}, I.~K. and {Baugh}, C.~M. and 
	{Bland-Hawthorn}, J. and {Bridges}, T. and {Cannon}, R. and 
	{Collins}, C. and {Couch}, W. and {Cross}, N. and {Dalton}, G. and 
	{De Propris}, R. and {Driver}, S.~P. and {Efstathiou}, G. and 
	{Ellis}, R.~S. and {Frenk}, C.~S. and {Glazebrook}, K. and {Lahav}, O. and 
	{Lewis}, I. and {Lumsden}, S. and {Maddox}, S. and {Madgwick}, D. and 
	{Sutherland}, W. and {Taylor}, K.},
    title = "{The 2dF Galaxy Redshift Survey: Final Data Release}",
  journal = {arXiv Astrophysics e-prints},
   eprint = {astro-ph/0306581},
     year = 2003,
    month = jun,
   adsurl = {https://ui.adsabs.harvard.edu/abs/2003astro.ph..6581C}
}

@ARTICLE{york2000sloan,
   author = {{York}, D.~G. and {Adelman}, J. and {Anderson}, Jr., J.~E. and 
	{Anderson}, S.~F. and {Annis}, J. and {Bahcall}, N.~A. and {Bakken}, J.~A. and 
	{Barkhouser}, R. and {Bastian}, S. and {Berman}, E. and {Boroski}, W.~N. and 
	{Bracker}, S. and {Briegel}, C. and {Briggs}, J.~W. and {Brinkmann}, J. and 
	{Brunner}, R. and {Burles}, S. and {Carey}, L. and {Carr}, M.~A. and 
	{Castander}, F.~J. and {Chen}, B. and {Colestock}, P.~L. and 
	{Connolly}, A.~J. and {Crocker}, J.~H. and {Csabai}, I. and 
	{Czarapata}, P.~C. and {Davis}, J.~E. and {Doi}, M. and {Dombeck}, T. and 
	{Eisenstein}, D. and {Ellman}, N. and {Elms}, B.~R. and {Evans}, M.~L. and 
	{Fan}, X. and {Federwitz}, G.~R. and {Fiscelli}, L. and {Friedman}, S. and 
	{Frieman}, J.~A. and {Fukugita}, M. and {Gillespie}, B. and 
	{Gunn}, J.~E. and {Gurbani}, V.~K. and {de Haas}, E. and {Haldeman}, M. and 
	{Harris}, F.~H. and {Hayes}, J. and {Heckman}, T.~M. and {Hennessy}, G.~S. and 
	{Hindsley}, R.~B. and {Holm}, S. and {Holmgren}, D.~J. and {Huang}, C.-h. and 
	{Hull}, C. and {Husby}, D. and {Ichikawa}, S.-I. and {Ichikawa}, T. and 
	{Ivezi{\'c}}, {\v Z}. and {Kent}, S. and {Kim}, R.~S.~J. and 
	{Kinney}, E. and {Klaene}, M. and {Kleinman}, A.~N. and {Kleinman}, S. and 
	{Knapp}, G.~R. and {Korienek}, J. and {Kron}, R.~G. and {Kunszt}, P.~Z. and 
	{Lamb}, D.~Q. and {Lee}, B. and {Leger}, R.~F. and {Limmongkol}, S. and 
	{Lindenmeyer}, C. and {Long}, D.~C. and {Loomis}, C. and {Loveday}, J. and 
	{Lucinio}, R. and {Lupton}, R.~H. and {MacKinnon}, B. and {Mannery}, E.~J. and 
	{Mantsch}, P.~M. and {Margon}, B. and {McGehee}, P. and {McKay}, T.~A. and 
	{Meiksin}, A. and {Merelli}, A. and {Monet}, D.~G. and {Munn}, J.~A. and 
	{Narayanan}, V.~K. and {Nash}, T. and {Neilsen}, E. and {Neswold}, R. and 
	{Newberg}, H.~J. and {Nichol}, R.~C. and {Nicinski}, T. and 
	{Nonino}, M. and {Okada}, N. and {Okamura}, S. and {Ostriker}, J.~P. and 
	{Owen}, R. and {Pauls}, A.~G. and {Peoples}, J. and {Peterson}, R.~L. and 
	{Petravick}, D. and {Pier}, J.~R. and {Pope}, A. and {Pordes}, R. and 
	{Prosapio}, A. and {Rechenmacher}, R. and {Quinn}, T.~R. and 
	{Richards}, G.~T. and {Richmond}, M.~W. and {Rivetta}, C.~H. and 
	{Rockosi}, C.~M. and {Ruthmansdorfer}, K. and {Sandford}, D. and 
	{Schlegel}, D.~J. and {Schneider}, D.~P. and {Sekiguchi}, M. and 
	{Sergey}, G. and {Shimasaku}, K. and {Siegmund}, W.~A. and {Smee}, S. and 
	{Smith}, J.~A. and {Snedden}, S. and {Stone}, R. and {Stoughton}, C. and 
	{Strauss}, M.~A. and {Stubbs}, C. and {SubbaRao}, M. and {Szalay}, A.~S. and 
	{Szapudi}, I. and {Szokoly}, G.~P. and {Thakar}, A.~R. and {Tremonti}, C. and 
	{Tucker}, D.~L. and {Uomoto}, A. and {Vanden Berk}, D. and {Vogeley}, M.~S. and 
	{Waddell}, P. and {Wang}, S.-i. and {Watanabe}, M. and {Weinberg}, D.~H. and 
	{Yanny}, B. and {Yasuda}, N. and {SDSS Collaboration}},
    title = "{The Sloan Digital Sky Survey: Technical Summary}",
  journal = {\aj},
   eprint = {astro-ph/0006396},
     year = 2000,
    month = sep,
   volume = 120,
    pages = {1579-1587},
      doi = {10.1086/301513},
   adsurl = {https://ui.adsabs.harvard.edu/abs/2000AJ....120.1579Y}
}

@ARTICLE{beutler20116df,
   author = {{Beutler}, F. and {Blake}, C. and {Colless}, M. and {Jones}, D.~H. and 
	{Staveley-Smith}, L. and {Poole}, G.~B. and {Campbell}, L. and 
	{Parker}, Q. and {Saunders}, W. and {Watson}, F.},
    title = "{The 6dF Galaxy Survey: z{\ap} 0 measurements of the growth rate and {$\sigma$}$_{8}$}",
  journal = {\mnras},
archivePrefix = "arXiv",
   eprint = {1204.4725},
     year = 2012,
    month = jul,
   volume = 423,
    pages = {3430-3444},
      doi = {10.1111/j.1365-2966.2012.21136.x},
   adsurl = {https://ui.adsabs.harvard.edu/abs/2012MNRAS.423.3430B}
}

@ARTICLE{blake2011wigglez,
   author = {{Blake}, C. and {Brough}, S. and {Colless}, M. and {Contreras}, C. and 
	{Couch}, W. and {Croom}, S. and {Davis}, T. and {Drinkwater}, M.~J. and 
	{Forster}, K. and {Gilbank}, D. and {Gladders}, M. and {Glazebrook}, K. and 
	{Jelliffe}, B. and {Jurek}, R.~J. and {Li}, I.-H. and {Madore}, B. and 
	{Martin}, D.~C. and {Pimbblet}, K. and {Poole}, G.~B. and {Pracy}, M. and 
	{Sharp}, R. and {Wisnioski}, E. and {Woods}, D. and {Wyder}, T.~K. and 
	{Yee}, H.~K.~C.},
    title = "{The WiggleZ Dark Energy Survey: the growth rate of cosmic structure since redshift z=0.9}",
  journal = {\mnras},
archivePrefix = "arXiv",
   eprint = {1104.2948},
     year = 2011,
    month = aug,
   volume = 415,
    pages = {2876-2891},
      doi = {10.1111/j.1365-2966.2011.18903.x},
   adsurl = {https://ui.adsabs.harvard.edu/abs/2011MNRAS.415.2876B}
}

@ARTICLE{blake2011wigglezb,
   author = {{Blake}, C. and {Glazebrook}, K. and {Davis}, T.~M. and {Brough}, S. and 
	{Colless}, M. and {Contreras}, C. and {Couch}, W. and {Croom}, S. and 
	{Drinkwater}, M.~J. and {Forster}, K. and {Gilbank}, D. and 
	{Gladders}, M. and {Jelliffe}, B. and {Jurek}, R.~J. and {Li}, I.-H. and 
	{Madore}, B. and {Martin}, D.~C. and {Pimbblet}, K. and {Poole}, G.~B. and 
	{Pracy}, M. and {Sharp}, R. and {Wisnioski}, E. and {Woods}, D. and 
	{Wyder}, T.~K. and {Yee}, H.~K.~C.},
    title = "{The WiggleZ Dark Energy Survey: measuring the cosmic expansion history using the Alcock-Paczynski test and distant supernovae}",
  journal = {\mnras},
archivePrefix = "arXiv",
   eprint = {1108.2637},
     year = 2011,
    month = dec,
   volume = 418,
    pages = {1725-1735},
      doi = {10.1111/j.1365-2966.2011.19606.x},
   adsurl = {https://ui.adsabs.harvard.edu/abs/2011MNRAS.418.1725B}
}

@ARTICLE{anderson2012clustering,
   author = {{Anderson}, L. and {Aubourg}, E. and {Bailey}, S. and {Bizyaev}, D. and 
	{Blanton}, M. and {Bolton}, A.~S. and {Brinkmann}, J. and {Brownstein}, J.~R. and 
	{Burden}, A. and {Cuesta}, A.~J. and {da Costa}, L.~A.~N. and 
	{Dawson}, K.~S. and {de Putter}, R. and {Eisenstein}, D.~J. and 
	{Gunn}, J.~E. and {Guo}, H. and {Hamilton}, J.-C. and {Harding}, P. and 
	{Ho}, S. and {Honscheid}, K. and {Kazin}, E. and {Kirkby}, D. and 
	{Kneib}, J.-P. and {Labatie}, A. and {Loomis}, C. and {Lupton}, R.~H. and 
	{Malanushenko}, E. and {Malanushenko}, V. and {Mandelbaum}, R. and 
	{Manera}, M. and {Maraston}, C. and {McBride}, C.~K. and {Mehta}, K.~T. and 
	{Mena}, O. and {Montesano}, F. and {Muna}, D. and {Nichol}, R.~C. and 
	{Nuza}, S.~E. and {Olmstead}, M.~D. and {Oravetz}, D. and {Padmanabhan}, N. and 
	{Palanque-Delabrouille}, N. and {Pan}, K. and {Parejko}, J. and 
	{P{\^a}ris}, I. and {Percival}, W.~J. and {Petitjean}, P. and 
	{Prada}, F. and {Reid}, B. and {Roe}, N.~A. and {Ross}, A.~J. and 
	{Ross}, N.~P. and {Samushia}, L. and {S{\'a}nchez}, A.~G. and 
	{Schlegel}, D.~J. and {Schneider}, D.~P. and {Sc{\'o}ccola}, C.~G. and 
	{Seo}, H.-J. and {Sheldon}, E.~S. and {Simmons}, A. and {Skibba}, R.~A. and 
	{Strauss}, M.~A. and {Swanson}, M.~E.~C. and {Thomas}, D. and 
	{Tinker}, J.~L. and {Tojeiro}, R. and {Maga{\~n}a}, M.~V. and 
	{Verde}, L. and {Wagner}, C. and {Wake}, D.~A. and {Weaver}, B.~A. and 
	{Weinberg}, D.~H. and {White}, M. and {Xu}, X. and {Y{\`e}che}, C. and 
	{Zehavi}, I. and {Zhao}, G.-B.},
    title = "{The clustering of galaxies in the SDSS-III Baryon Oscillation Spectroscopic Survey: baryon acoustic oscillations in the Data Release 9 spectroscopic galaxy sample}",
  journal = {\mnras},
archivePrefix = "arXiv",
   eprint = {1203.6594},
     year = 2012,
    month = dec,
   volume = 427,
    pages = {3435-3467},
      doi = {10.1111/j.1365-2966.2012.22066.x},
   adsurl = {https://ui.adsabs.harvard.edu/abs/2012MNRAS.427.3435A}
}

@ARTICLE{alam2017clustering,
   author = {{Alam}, S. and {Ata}, M. and {Bailey}, S. and {Beutler}, F. and 
	{Bizyaev}, D. and {Blazek}, J.~A. and {Bolton}, A.~S. and {Brownstein}, J.~R. and 
	{Burden}, A. and {Chuang}, C.-H. and {Comparat}, J. and {Cuesta}, A.~J. and 
	{Dawson}, K.~S. and {Eisenstein}, D.~J. and {Escoffier}, S. and 
	{Gil-Mar{\'{\i}}n}, H. and {Grieb}, J.~N. and {Hand}, N. and 
	{Ho}, S. and {Kinemuchi}, K. and {Kirkby}, D. and {Kitaura}, F. and 
	{Malanushenko}, E. and {Malanushenko}, V. and {Maraston}, C. and 
	{McBride}, C.~K. and {Nichol}, R.~C. and {Olmstead}, M.~D. and 
	{Oravetz}, D. and {Padmanabhan}, N. and {Palanque-Delabrouille}, N. and 
	{Pan}, K. and {Pellejero-Ibanez}, M. and {Percival}, W.~J. and 
	{Petitjean}, P. and {Prada}, F. and {Price-Whelan}, A.~M. and 
	{Reid}, B.~A. and {Rodr{\'{\i}}guez-Torres}, S.~A. and {Roe}, N.~A. and 
	{Ross}, A.~J. and {Ross}, N.~P. and {Rossi}, G. and {Rubi{\~n}o-Mart{\'{\i}}n}, J.~A. and 
	{Saito}, S. and {Salazar-Albornoz}, S. and {Samushia}, L. and 
	{S{\'a}nchez}, A.~G. and {Satpathy}, S. and {Schlegel}, D.~J. and 
	{Schneider}, D.~P. and {Sc{\'o}ccola}, C.~G. and {Seo}, H.-J. and 
	{Sheldon}, E.~S. and {Simmons}, A. and {Slosar}, A. and {Strauss}, M.~A. and 
	{Swanson}, M.~E.~C. and {Thomas}, D. and {Tinker}, J.~L. and 
	{Tojeiro}, R. and {Maga{\~n}a}, M.~V. and {Vazquez}, J.~A. and 
	{Verde}, L. and {Wake}, D.~A. and {Wang}, Y. and {Weinberg}, D.~H. and 
	{White}, M. and {Wood-Vasey}, W.~M. and {Y{\`e}che}, C. and 
	{Zehavi}, I. and {Zhai}, Z. and {Zhao}, G.-B.},
    title = "{The clustering of galaxies in the completed SDSS-III Baryon Oscillation Spectroscopic Survey: cosmological analysis of the DR12 galaxy sample}",
  journal = {\mnras},
archivePrefix = "arXiv",
   eprint = {1607.03155},
     year = 2017,
    month = sep,
   volume = 470,
    pages = {2617-2652},
      doi = {10.1093/mnras/stx721},
   adsurl = {https://ui.adsabs.harvard.edu/abs/2017MNRAS.470.2617A}
}

@ARTICLE{LI15,
   author = {{Li}, X.-D. and {Park}, C. and {Sabiu}, C.~G. and {Kim}, J.},
    title = "{Cosmological constraints from the redshift dependence of the Alcock-Paczynski test and volume effect: galaxy two-point correlation function}",
  journal = {\mnras},
archivePrefix = "arXiv",
   eprint = {1504.00740},
     year = 2015,
    month = jun,
   volume = 450,
    pages = {807-814},
      doi = {10.1093/mnras/stv622},
   adsurl = {https://ui.adsabs.harvard.edu/abs/2015MNRAS.450..807L}
}

@ARTICLE{Sheth:2004vb,
       author = {{Sheth}, Ravi K. and {Tormen}, Giuseppe},
        title = "{On the environmental dependence of halo formation}",
      journal = {\mnras},
         year = 2004,
        month = jun,
       volume = {350},
       number = {4},
        pages = {1385-1390},
          doi = {10.1111/j.1365-2966.2004.07733.x},
archivePrefix = {arXiv},
       eprint = {astro-ph/0402237},
 primaryClass = {astro-ph},
       adsurl = {https://ui.adsabs.harvard.edu/abs/2004MNRAS.350.1385S}
}

@ARTICLE{Satpathy:2019nvo,
       author = {{Satpathy}, Siddharth and {A C Croft}, Rupert and {Ho}, Shirley and
         {Li}, Baojiu},
        title = "{Measurement of marked correlation functions in SDSS-III Baryon Oscillation Spectroscopic Survey using LOWZ galaxies in Data Release 12}",
      journal = {\mnras},
         year = 2019,
        month = apr,
       volume = {484},
       number = {2},
        pages = {2148-2165},
          doi = {10.1093/mnras/stz009},
archivePrefix = {arXiv},
       eprint = {1901.01447},
 primaryClass = {astro-ph.CO},
       adsurl = {https://ui.adsabs.harvard.edu/abs/2019MNRAS.484.2148S}
}

@ARTICLE{Park:2019mvn,
       author = {{Park}, Hyunbae and {Park}, Changbom and {Sabiu}, Cristiano G. and
         {Li}, Xiao-dong and {Hong}, Sungwook E. and {Kim}, Juhan and
         {Tonegawa}, Motonari and {Zheng}, Yi},
        title = "{Alcock-Paczynski Test with the Evolution of Redshift-space Galaxy Clustering Anisotropy}",
      journal = {\apj},
         year = 2019,
        month = aug,
       volume = {881},
       number = {2},
          eid = {146},
        pages = {146},
          doi = {10.3847/1538-4357/ab2da1},
archivePrefix = {arXiv},
       eprint = {1904.05503},
 primaryClass = {astro-ph.CO},
       adsurl = {https://ui.adsabs.harvard.edu/abs/2019ApJ...881..146P}
}

@ARTICLE{Ma_2020,
       author = {{Ma}, Qinglin and {Guo}, Yiqing and {Li}, Xiao-Dong and {Wang}, Xin and
         {Miao}, Haitao and {Li}, Zhigang and {Sabiu}, Cristiano G. and
         {Park}, Hyunbae},
        title = "{Cosmological Constraints from the Redshift Dependence of the Alcock--Paczynski Effect: Possibility of Estimating the Nonlinear Systematics Using Fast Simulations}",
      journal = {\apj},
         year = 2020,
        month = feb,
       volume = {890},
       number = {2},
          eid = {92},
        pages = {92},
          doi = {10.3847/1538-4357/ab6aa3},
archivePrefix = {arXiv},
       eprint = {1908.10595},
 primaryClass = {astro-ph.CO},
       adsurl = {https://ui.adsabs.harvard.edu/abs/2020ApJ...890...92M}
}

@ARTICLE{White_2009,
       author = {{White}, Martin and {Padmanabhan}, Nikhil},
        title = "{Breaking halo occupation degeneracies with marked statistics}",
      journal = {\mnras},
         year = 2009,
        month = jun,
       volume = {395},
       number = {4},
        pages = {2381-2384},
          doi = {10.1111/j.1365-2966.2009.14732.x},
archivePrefix = {arXiv},
       eprint = {0812.4288},
 primaryClass = {astro-ph},
       adsurl = {https://ui.adsabs.harvard.edu/abs/2009MNRAS.395.2381W}
}

@ARTICLE{Lucy1977,
       author = {{Lucy}, L.~B.},
        title = "{A numerical approach to the testing of the fission hypothesis.}",
      journal = {\aj},
         year = 1977,
        month = dec,
       volume = {82},
        pages = {1013-1024},
          doi = {10.1086/112164},
       adsurl = {https://ui.adsabs.harvard.edu/abs/1977AJ.....82.1013L}
}

@ARTICLE{Gong_2019,
       author = {{Gong}, Yan and {Liu}, Xiangkun and {Cao}, Ye and {Chen}, Xuelei and
         {Fan}, Zuhui and {Li}, Ran and {Li}, Xiao-Dong and {Li}, Zhigang and
         {Zhang}, Xin and {Zhan}, Hu},
        title = "{Cosmology from the Chinese Space Station Optical Survey (CSS-OS)}",
      journal = {\apj},
         year = 2019,
        month = oct,
       volume = {883},
       number = {2},
          eid = {203},
        pages = {203},
          doi = {10.3847/1538-4357/ab391e},
archivePrefix = {arXiv},
       eprint = {1901.04634},
 primaryClass = {astro-ph.CO},
       adsurl = {https://ui.adsabs.harvard.edu/abs/2019ApJ...883..203G}
}

@ARTICLE{Beisbart2000,
       author = {{Beisbart}, Claus and {Kerscher}, Martin},
        title = "{Luminosity- and Morphology-dependent Clustering of Galaxies}",
      journal = {\apj},
         year = 2000,
        month = dec,
       volume = {545},
       number = {1},
        pages = {6-25},
          doi = {10.1086/317788},
archivePrefix = {arXiv},
       eprint = {astro-ph/0003358},
 primaryClass = {astro-ph},
       adsurl = {https://ui.adsabs.harvard.edu/abs/2000ApJ...545....6B}
}

@INBOOK{Beisbart2002,
       author = {{Beisbart}, Claus and {Kerscher}, Martin and {Mecke}, Klaus},
        title = "{Mark Correlations: Relating Physical Properties to Spatial Distributions}",
    booktitle = {Morphology of Condensed Matter},
         year = 2002,
       editor = {{Mecke}, K. and {Stoyan}, D.},
       volume = {600},
        pages = {358-390},
       adsurl = {https://ui.adsabs.harvard.edu/abs/2002LNP...600..358B}
}

@ARTICLE{Gottl2002,
       author = {{Gottl{\"o}ber}, S. and {Kerscher}, M. and {Kravtsov}, A.~V. and
         {Faltenbacher}, A. and {Klypin}, A. and {M{\"u}ller}, V.},
        title = "{Spatial distribution of galactic halos and their merger histories}",
      journal = {Astronomy \& Astrophysics },
         year = 2002,
        month = jun,
       volume = {387},
        pages = {778-787},
          doi = {10.1051/0004-6361:20020339},
archivePrefix = {arXiv},
       eprint = {astro-ph/0203148},
 primaryClass = {astro-ph},
       adsurl = {https://ui.adsabs.harvard.edu/abs/2002A&A...387..778G}
}

@ARTICLE{Skibba2006,
       author = {{Skibba}, Ramin and {Sheth}, Ravi K. and {Connolly}, Andrew J. and
         {Scranton}, Ryan},
        title = "{The luminosity-weighted or `marked' correlation function}",
      journal = {\mnras},
         year = 2006,
        month = jun,
       volume = {369},
       number = {1},
        pages = {68-76},
          doi = {10.1111/j.1365-2966.2006.10196.x},
archivePrefix = {arXiv},
       eprint = {astro-ph/0512463},
 primaryClass = {astro-ph},
       adsurl = {https://ui.adsabs.harvard.edu/abs/2006MNRAS.369...68S}
}

@ARTICLE{Gingold1977,
       author = {{Gingold}, R.~A. and {Monaghan}, J.~J.},
        title = "{Smoothed particle hydrodynamics: theory and application to non-spherical stars.}",
      journal = {\mnras},
         year = 1977,
        month = nov,
       volume = {181},
        pages = {375-389},
          doi = {10.1093/mnras/181.3.375},
       adsurl = {https://ui.adsabs.harvard.edu/abs/1977MNRAS.181..375G}
}

@article{White_2016,
	doi = {10.1088/1475-7516/2016/11/057},
	url = {https://doi.org/10.1088%2F1475-7516%2F2016%2F11%2F057},
	year = 2016,
	month = {nov},
	publisher = {{IOP} Publishing},
	volume = {2016},
	number = {11},
	pages = {057--057},
	author = {Martin White},
	title = {A marked correlation function for constraining modified gravity models},
	journal = {Journal of Cosmology and Astroparticle Physics},
}

@article{PMS2020,
    author = "Philcox, Oliver H. E. and Massara, Elena and Spergel, David N.",
    title = "{What does the marked power spectrum measure? Insights from perturbation theory}",
    eprint = "2006.10055",
    archivePrefix = "arXiv",
    primaryClass = "astro-ph.CO",
    doi = "10.1103/PhysRevD.102.043516",
    journal = "Phys. Rev. D",
    volume = "102",
    number = "4",
    pages = "043516",
    year = "2020"
}

@ARTICLE{2020arXiv200111024M,
       author = {{Massara}, Elena and {Villaescusa-Navarro}, Francisco and {Ho}, Shirley and
         {Dalal}, Neal and {Spergel}, David N.},
        title = "{Using the Marked Power Spectrum to Detect the Signature of Neutrinos in Large-Scale Structure}",
      journal = {arXiv e-prints},
         year = 2020,
        month = jan,
          eid = {arXiv:2001.11024},
        pages = {arXiv:2001.11024},
archivePrefix = {arXiv},
       eprint = {2001.11024},
 primaryClass = {astro-ph.CO},
       adsurl = {https://ui.adsabs.harvard.edu/abs/2020arXiv200111024M}
}

@misc{Sheth:2005aj,
      title={Marked correlations in galaxy formation models}, 
      author={Ravi K. Sheth and Andrew J. Connolly and Ramin Skibba},
      year={2005},
      eprint={astro-ph/0511773},
      archivePrefix={arXiv},
      primaryClass={astro-ph},
      url={https://arxiv.org/abs/astro-ph/0511773}, 
}

@article{Xiao_2022,
    author = {Xiao, Liang and Huang, Zhiqi and Zheng, Yi and Wang, Xin and Li, Xiao-Dong},
    title = {Tomographic Alcock–Paczynski method with redshift errors},
    journal = {Monthly Notices of the Royal Astronomical Society},
    volume = {518},
    number = {4},
    pages = {6253-6261},
    year = {2022},
    month = {12},
    issn = {0035-8711},
    doi = {10.1093/mnras/stac2996},
    url = {https://doi.org/10.1093/mnras/stac2996},
    eprint = {https://academic.oup.com/mnras/article-pdf/518/4/6253/48012059/stac2996.pdf},
}

@article{XiaoYuan_2022,
    author = {Xiao, Xiaoyuan and Yang, Yizhao and Luo, Xiaolin and Ding, Jiacheng and Huang, Zhiqi and Wang, Xin and Zheng, Yi and Sabiu, Cristiano G and Forero-Romero, Jaime and Miao, Haitao and Li, Xiao-Dong},
    title = {Cosmological constraints from the density gradient weighted correlation function},
    journal = {Monthly Notices of the Royal Astronomical Society},
    volume = {513},
    number = {1},
    pages = {595-603},
    year = {2022},
    month = {04},
    issn = {0035-8711},
    doi = {10.1093/mnras/stac879},
    url = {https://doi.org/10.1093/mnras/stac879},
    eprint = {https://academic.oup.com/mnras/article-pdf/513/1/595/43426298/stac879.pdf},
}

@article{Yang:2020ysv,
    author = "Yang, Yizhao and others",
    title = "{Using the Mark Weighted Correlation Functions to Improve the Constraints on Cosmological Parameters}",
    eprint = "2007.03150",
    archivePrefix = "arXiv",
    primaryClass = "astro-ph.CO",
    doi = "10.3847/1538-4357/aba35b",
    journal = "Astrophys. J.",
    volume = "900",
    number = "1",
    pages = "6",
    year = "2020"
}

@article{Dong_2023,
doi = {10.3847/1538-4357/acd185},
url = {https://dx.doi.org/10.3847/1538-4357/acd185},
year = {2023},
month = {aug},
publisher = {The American Astronomical Society},
volume = {953},
number = {1},
pages = {98},
author = {Dong, Fuyu and Park, Changbom and Hong, Sungwook E. and Kim, Juhan and Hwang, Ho Seong and Park, Hyunbae and Appleby, Stephen},
title = {Tomographic Alcock--Paczyński Test with Redshift-space Correlation Function: Evidence for the Dark Energy Equation-of-state Parameter $w > -1$},
journal = {The Astrophysical Journal}
}

@article{Crittenden_2009,
    author = "Crittenden, Robert G. and Pogosian, Levon and Zhao, Gong-Bo",
    title = "{Investigating dark energy experiments with principal components}",
    eprint = "astro-ph/0510293",
    archivePrefix = "arXiv",
    doi = "10.1088/1475-7516/2009/12/025",
    journal = "JCAP",
    volume = "12",
    pages = "025",
    year = "2009"
}

@article{Huterer_2003,
    author = "Huterer, Dragan and Starkman, Glenn",
    title = "{Parameterization of dark-energy properties: A Principal-component approach}",
    eprint = "astro-ph/0207517",
    archivePrefix = "arXiv",
    reportNumber = "CWRU-07-02",
    doi = "10.1103/PhysRevLett.90.031301",
    journal = "Phys. Rev. Lett.",
    volume = "90",
    pages = "031301",
    year = "2003"
}

@article{emcee,
   author = {{Foreman-Mackey}, D. and {Hogg}, D.~W. and {Lang}, D. and {Goodman}, J.},
    title = {emcee: The MCMC Hammer},
  journal = {PASP},
     year = 2013,
   volume = 125,
    pages = {306-312},
   eprint = {1202.3665},
      doi = {10.1086/670067}
}

@article{Yin:2024iqp,
    author = "Yin, Fenfen and Ding, Jiacheng and Lai, Limin and Zhang, Wei and Xiao, Liang and Wang, Zihan and Forero-Romero, Jaime and Zhang, Le and Li, Xiao-Dong",
    title = "{Improving SDSS cosmological constraints through \ensuremath{\beta}-skeleton weighted correlation functions}",
    eprint = "2403.14165",
    archivePrefix = "arXiv",
    primaryClass = "astro-ph.CO",
    doi = "10.1103/PhysRevD.109.123537",
    journal = "Phys. Rev. D",
    volume = "109",
    number = "12",
    pages = "123537",
    year = "2024"
}

@article{Lai:2023dzp,
    author = "Lai, Limin and others",
    title = "{Improving constraint on \ensuremath{\Omega}$_{m}$ from SDSS using marked correlation functions}",
    eprint = "2312.03244",
    archivePrefix = "arXiv",
    primaryClass = "astro-ph.CO",
    doi = "10.1007/s11433-023-2384-4",
    journal = "Sci. China Phys. Mech. Astron.",
    volume = "67",
    number = "8",
    pages = "289512",
    year = "2024"
}

@article{stoyan_correlations_1984,
	title = {On Correlations of Marked Point Processes},
	volume = {116},
	issn = {1522-2616},
	url = {https://onlinelibrary.wiley.com/doi/abs/10.1002/mana.19841160115},
	doi = {10.1002/mana.19841160115},
	number = {1},
	urldate = {2025-04-29},
	journal = {Mathematische Nachrichten},
	author = {Stoyan, Dietrich},
	year = {1984},
	pages = {197--207},
}

@article{Massara:2022zrf,
    author = "Massara, Elena and Villaescusa-Navarro, Francisco and Hahn, ChangHoon and Abidi, Muntazir M. and Eickenberg, Michael and Ho, Shirley and Lemos, Pablo and Moradinezhad Dizgah, Azadeh and Blancard, Bruno R\'egaldo-Saint",
    title = "{Cosmological Information in the Marked Power Spectrum of the Galaxy Field}",
    eprint = "2206.01709",
    archivePrefix = "arXiv",
    primaryClass = "astro-ph.CO",
    doi = "10.3847/1538-4357/acd44d",
    journal = "Astrophys. J.",
    volume = "951",
    number = "1",
    pages = "70",
    year = "2023"
}

@article{Jung:2024esv,
    author = "Jung, Gabriel and Ravenni, Andrea and Liguori, Michele and Baldi, Marco and Coulton, William R. and Villaescusa-Navarro, Francisco and Wandelt, Benjamin D.",
    title = "{Quijote-PNG: Optimizing the Summary Statistics to Measure Primordial Non-Gaussianity}",
    eprint = "2403.00490",
    archivePrefix = "arXiv",
    primaryClass = "astro-ph.CO",
    doi = "10.3847/1538-4357/ad83bd",
    journal = "Astrophys. J.",
    volume = "976",
    number = "1",
    pages = "109",
    year = "2024"
}

@article{Philcox:2020fqx,
    author = "Philcox, Oliver H. E. and Massara, Elena and Spergel, David N.",
    title = "{What does the marked power spectrum measure? Insights from perturbation theory}",
    eprint = "2006.10055",
    archivePrefix = "arXiv",
    primaryClass = "astro-ph.CO",
    doi = "10.1103/PhysRevD.102.043516",
    journal = "Phys. Rev. D",
    volume = "102",
    number = "4",
    pages = "043516",
    year = "2020"
}

@article{Philcox:2020srd,
    author = "Philcox, Oliver H. E. and Aviles, Alejandro and Massara, Elena",
    title = "{Modeling the Marked Spectrum of Matter and Biased Tracers in Real- and Redshift-Space}",
    eprint = "2010.05914",
    archivePrefix = "arXiv",
    primaryClass = "astro-ph.CO",
    doi = "10.1088/1475-7516/2021/03/038",
    journal = "JCAP",
    volume = "03",
    pages = "038",
    year = "2021"
}

@article{Ebina:2024zkv,
    author = "Ebina, Haruki and White, Martin",
    title = "{An analytically tractable marked power spectrum}",
    eprint = "2409.17133",
    archivePrefix = "arXiv",
    primaryClass = "astro-ph.CO",
    doi = "10.1088/1475-7516/2025/01/150",
    journal = "JCAP",
    volume = "01",
    pages = "150",
    year = "2025"
}

@article{Massara:2024cvu,
    author = "Massara, Elena and Hahn, ChangHoon and Eickenberg, Michael and Ho, Shirley and Hou, Jiamin and Lemos, Pablo and Modi, Chirag and Moradinezhad Dizgah, Azadeh and Parker, Liam and Blancard, Bruno R\'egaldo-Saint",
    title = "{{\textbackslash{}sc SimBIG}: Cosmological Constraints using Simulation-Based Inference of Galaxy Clustering with Marked Power Spectra}",
    eprint = "2404.04228",
    journal = {arXiv e-prints},
    archivePrefix = "arXiv",
    primaryClass = "astro-ph.CO",
    month = "4",
    year = "2024"
}

@article{scikit-learn,
  title={Scikit-learn: Machine Learning in {P}ython},
  author={Pedregosa, F. and Varoquaux, G. and Gramfort, A. and Michel, V.
          and Thirion, B. and Grisel, O. and Blondel, M. and Prettenhofer, P.
          and Weiss, R. and Dubourg, V. and Vanderplas, J. and Passos, A. and
          Cournapeau, D. and Brucher, M. and Perrot, M. and Duchesnay, E.},
  journal={Journal of Machine Learning Research},
  volume={12},
  pages={2825--2830},
  year={2011}
}

@article{MOPED,
    author = {Heavens, Alan F. and Jimenez, Raul and Lahav, Ofer},
    title = {Massive lossless data compression and multiple parameter estimation from galaxy spectra},
    journal = {Monthly Notices of the Royal Astronomical Society},
    volume = {317},
    number = {4},
    pages = {965-972},
    year = {2000},
    month = {10},
    issn = {0035-8711},
    doi = {10.1046/j.1365-8711.2000.03692.x},
    url = {https://doi.org/10.1046/j.1365-8711.2000.03692.x},
    eprint = {https://academic.oup.com/mnras/article-pdf/317/4/965/3505248/317-4-965.pdf},
}

@article{Miao_2022,
    author = {Miao, Haitao and Gong, Yan and Chen, Xuelei and Huang, Zhiqi and Li, Xiao-Dong and Zhan, Hu},
    title = {Cosmological constraint precision of photometric and spectroscopic multi-probe surveys of China Space Station Telescope (CSST)},
    journal = {Monthly Notices of the Royal Astronomical Society},
    volume = {519},
    number = {1},
    pages = {1132-1148},
    year = {2022},
    month = {12},
    issn = {0035-8711},
    doi = {10.1093/mnras/stac3583},
    url = {https://doi.org/10.1093/mnras/stac3583},
    eprint = {https://academic.oup.com/mnras/article-pdf/519/1/1132/48375772/stac3583.pdf},
}

@ARTICLE{CLPT,
       author = {{Carlson}, Jordan and {Reid}, Beth and {White}, Martin},
        title = "{Convolution Lagrangian perturbation theory for biased tracers}",
      journal = {Monthly Notices of the Royal Astronomical Society},
         year = 2013,
        month = feb,
       volume = {429},
       number = {2},
        pages = {1674-1685},
          doi = {10.1093/mnras/sts457},
archivePrefix = {arXiv},
       eprint = {1209.0780},
 primaryClass = {astro-ph.CO},
       adsurl = {https://ui.adsabs.harvard.edu/abs/2013MNRAS.429.1674C}
}

@ARTICLE{GS_model,
       author = {{Reid}, Beth A. and {White}, Martin},
        title = "{Towards an accurate model of the redshift-space clustering of haloes in the quasi-linear regime}",
      journal = {Monthly Notices of the Royal Astronomical Society},
         year = 2011,
        month = nov,
       volume = {417},
       number = {3},
        pages = {1913-1927},
          doi = {10.1111/j.1365-2966.2011.19379.x},
archivePrefix = {arXiv},
       eprint = {1105.4165},
 primaryClass = {astro-ph.CO},
       adsurl = {https://ui.adsabs.harvard.edu/abs/2011MNRAS.417.1913R}
}

@ARTICLE{CLPT_GSRSD,
       author = {{Wang}, Lile and {Reid}, Beth and {White}, Martin},
        title = "{An analytic model for redshift-space distortions}",
      journal = {Monthly Notices of the Royal Astronomical Society},
         year = 2014,
        month = jan,
       volume = {437},
       number = {1},
        pages = {588-599},
          doi = {10.1093/mnras/stt1916},
archivePrefix = {arXiv},
       eprint = {1306.1804},
 primaryClass = {astro-ph.CO},
       adsurl = {https://ui.adsabs.harvard.edu/abs/2014MNRAS.437..588W}
}

@article{EFT1,
   title={Effective field theory of cosmic acceleration: An implementation in CAMB},
   volume={89},
   ISSN={1550-2368},
   url={http://dx.doi.org/10.1103/PhysRevD.89.103530},
   DOI={10.1103/physrevd.89.103530},
   number={10},
   journal={Physical Review D},
   publisher={American Physical Society (APS)},
   author={Hu, Bin and Raveri, Marco and Frusciante, Noemi and Silvestri, Alessandra},
   year={2014},
   month=may }

@article{EFT2,
   title={Dark energy or modified gravity? An effective field theory approach},
   volume={2013},
   ISSN={1475-7516},
   url={http://dx.doi.org/10.1088/1475-7516/2013/08/010},
   DOI={10.1088/1475-7516/2013/08/010},
   number={08},
   journal={Journal of Cosmology and Astroparticle Physics},
   publisher={IOP Publishing},
   author={Bloomfield, Jolyon and Flanagan, Eanna E and Park, Minjoon and Watson, Scott},
   year={2013},
   month=aug, pages={010–010} }

@article{EFT3,
   title={Effective field theory of cosmological perturbations},
   volume={30},
   ISSN={1361-6382},
   url={http://dx.doi.org/10.1088/0264-9381/30/21/214007},
   DOI={10.1088/0264-9381/30/21/214007},
   number={21},
   journal={Classical and Quantum Gravity},
   publisher={IOP Publishing},
   author={Piazza, Federico and Vernizzi, Filippo},
   year={2013},
   month=oct, pages={214007} }

@article{COLA,
   title={Solving large scale structure in ten easy steps with COLA},
   volume={2013},
   ISSN={1475-7516},
   url={http://dx.doi.org/10.1088/1475-7516/2013/06/036},
   DOI={10.1088/1475-7516/2013/06/036},
   number={06},
   journal={Journal of Cosmology and Astroparticle Physics},
   publisher={IOP Publishing},
   author={Tassev, Svetlin and Zaldarriaga, Matias and Eisenstein, Daniel J},
   year={2013},
   month=jun, pages={036–036} }

@article{mg_COLA,
   title={Fast route to nonlinear clustering statistics in modified gravity theories},
   volume={91},
   ISSN={1550-2368},
   url={http://dx.doi.org/10.1103/PhysRevD.91.123507},
   DOI={10.1103/physrevd.91.123507},
   number={12},
   journal={Physical Review D},
   publisher={American Physical Society (APS)},
   author={Winther, Hans A. and Ferreira, Pedro G.},
   year={2015},
   month=jun }

\appendix

\section{Some tests about insignificant cosmological dependence of redshift evolution of RSD} \label{sec:appendix_RSD}

It is a core assumption that the cosmological dependence of redshift evolution of RSD is slight, and therefore we can ignore it and just estimate the systematic correction in the fiducial cosmology. In this section, we will provide some tests related to this assumption.

We first consider it using a physical model for the $\xi(s, \mu;z)$, based on perturbation theory (CLPT\citep{CLPT}, EFT\citep{EFT1, EFT2, EFT3}, etc.). We adopt the CLPT (Convolution Lagrangian Perturbation Theory), which is a new formulation of Lagrangian perturbation theory that allows accurate predictions of the real space and redshift space correlation functions of the dark matter haloes, and we adopt the Gaussian streaming model \citep{GS_model} to create the redshift space statistics. The code we adopt is  CLPT$\_$GSRSD\footnote{https://github.com/wll745881210/CLPT$\_$GSRSD}\citep{CLPT_GSRSD}.  We fit the first and second order Lagrangian bias parameters: $\langle F^{\prime} \rangle$ and $\langle F^{\prime \prime} \rangle$ with the simulation we use in this work (\texttt{BIGMD}) and then calculate $\xi(s, \mu;z_1)$ and $\xi(s, \mu;z_2)$ with RSD. 

We adopt three other cosmologies: one is $\Lambda\mathbf{CDM}$ with $\Omega_m = 0.35$ (case I), while the others are $w\mathrm{CDM}$ with $\Omega_m=0.3071, w=-0.7$ (case II) and $\Omega_m=0.45, w=-1.3$ (case III), and other cosmological parameters are the same as BIGMD used in this work. We found the obvious cosmological dependence of the redshift evolution of RSD with integration of $6 \le s \le 40 \mathrm{Mpc}/h$, especially the cases with different $\Omega_m$, however, it cannot be seen if choosing $20-60 \mathrm{Mpc}/h$. The results are shown in Fig.\ref{fig:CLPT}. We believe that this cannot explain that our assumption is invalid at $6 \le s \le 40 \mathrm{Mpc}/h$ because it shows a significant deviation from the simulation used for the fit.

To confirm this, we then conduct additional simulations based on a fast algorithm: COLA(COmoving Lagrangian Acceleration\citep{COLA, mg_COLA}), which is an N-body method for solving for Large Scale Structure (LSS) in a frame that is comoving with observers following trajectories calculated in Lagrangian Perturbation Theory (LPT). It can straightforwardly trade accuracy at small-scales in order to gain computational speed without sacrificing accuracy at large scales, and according to the tests of \cite{Ma_2020}, it meets our requirements in this part. 

The adopted cosmologies are the same as those in the test of CLPT, with a box with $L = 800 \mathrm{Mpc}/h$ and $1280^3$ DM particles with the PM grid size $(1280 \times 3)^3$ to calculate the force. We utilize \texttt{ROCKSTAR} halo finder to find halos and subhalos and maintain a constant number density $\overline{n} = 0.001 (h /\mathrm{Mpc})^3$ in all snapshots. Based on it, we calculate $\xi(s,\mu;z_1)$ and $\xi(s,\mu;z_2)$ with Landy–Szalay estimator(Eq.\ref{eq:xi_primary}) and then calculate the anisotropic redshift evolution of RSD. The results are shown in Fig.\ref{fig:COLA}. In each cosmologies we ran 10 realizations to suppress the randomness. The evolution of RSD in the fiducial case is consistent with that in BIGMD. In addition, we found that the cosmological dependence of the redshift evolution of RSD is indeed slight in the sense of simulation, especially compared with the AP effects.

In summary, the insignificant cosmological dependence of redshift evolution of RSD is tested in the sense of simulation. Besides, it's difficult to be tested through perturbation theory because the scale (6-40$\mathrm{Mpc}/h$) in our method is deeply affected by RSD, which is difficult to estimate accurately within this scale in the sense of perturbation theory.

\begin{figure*}[!htpb]
    \centering
    \includegraphics[width=0.9 \textwidth]{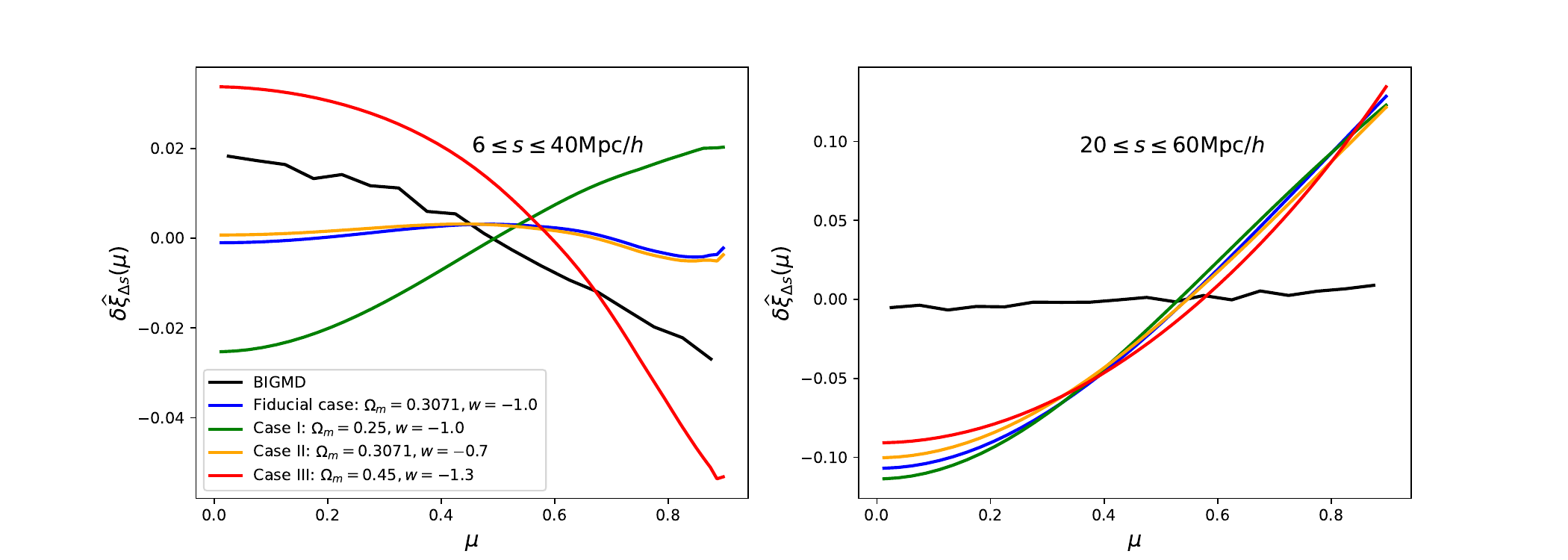}
    \caption{The anisotropic redshift evolution of RSD in the cosmologies of fiducial case($\Omega_m = 0.3071, w = -1.0$), case I($\Omega_m = 0.25, w = -1.0$), case II($\Omega_m = 0.3071, w = -0.7$) and case III($\Omega_m = 0.45, w = -1.3$), with the integration of $6 \leq s \leq 40 \mathrm{Mpc}/h$  (the left panel) and $20 \leq s \leq 60 \mathrm{Mpc}/h$ (the right panel), respectively. Besides, we exclude $\mu > 0.9$ to avoid effects of FOG. The four cases in this test are denoted by lines with different colors. Besides, the anisotropic redshift evolution of RSD in BIGMD is also attached with the black lines. }
    \label{fig:CLPT}
\end{figure*}

\begin{figure*}[!htpb]
    \centering
    \includegraphics[width=0.9 \textwidth]{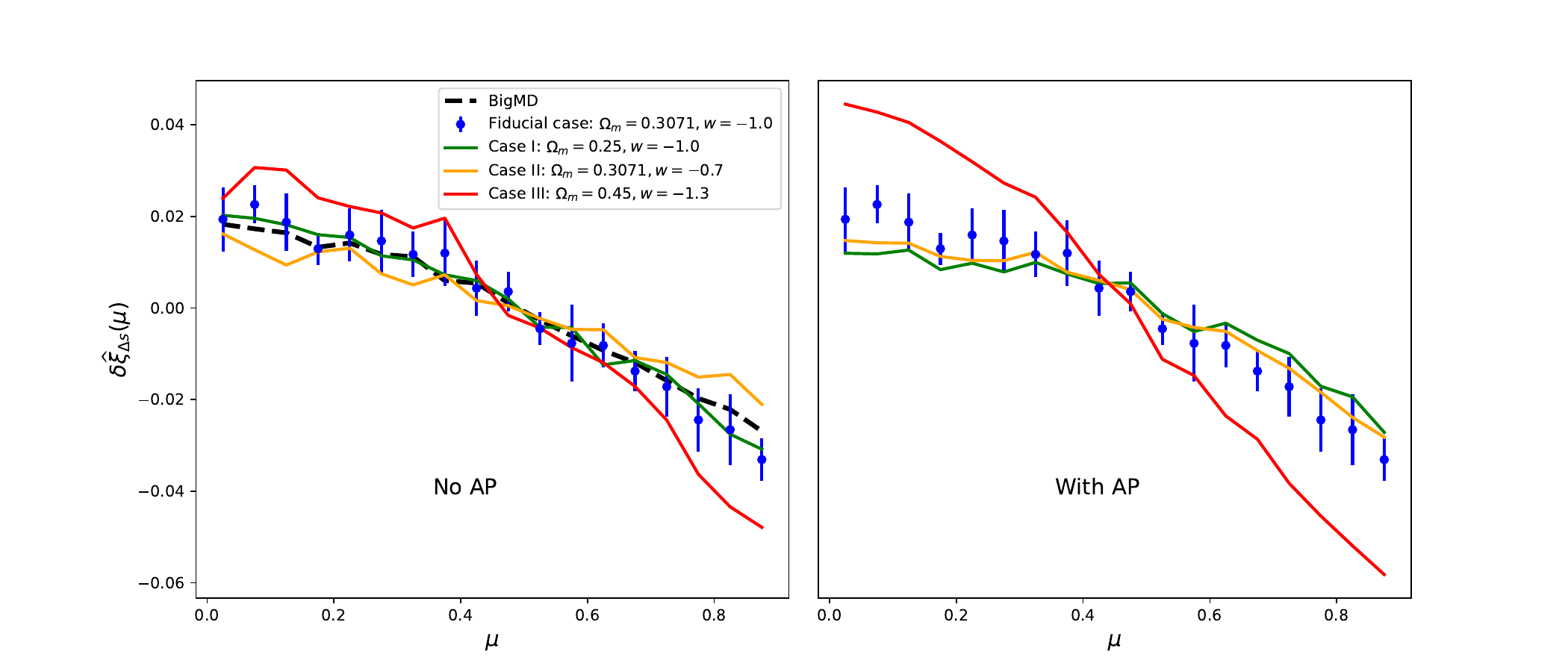}
    \caption{The anisotropic redshift evolution of RSD (the left panel) and AP signals (the right panel) in the four cosmologies: fiducial case, case I, case II and case III, which are the same as Fig.\ref{fig:CLPT}, but now we obtain the results with COLA. We integrate $\xi(s,\mu)$ with $s$ in $6 \leq s \leq 40 \mathrm{Mpc}/h$ and exclude $\mu > 0.9$ to avoid effects of FOG. Each case we ran 10 realizations to reduce the randomness, but we only plot errorbars in the fiducial case for convenience to view. The evolution of RSD in the fiducial case is consistent with that in BIGMD, which is denoted by black dashed line on the left panel. The deviation between fiducial cosmology and other comsologies on the right panel replects the cosmological evolution of the redshift evolution of the mixture of RSD and pure AP signals instead of just the pure AP signals, however, it is still significant that the cosmological dependence of the redshift evolution of RSD is much slighter than that of the pure AP signals.}
    \label{fig:COLA}
\end{figure*}

\section{Robustness check about the covariance estimation}\label{sec:appendix_cov}

Here we do a simple test to see whether the results are insensitive to covariance estimation \footnote{Note that our PCA models are trained from catalogues having differnt AP effects (i.e. changing the background of the mocks to different sets of wrongly induced cosmological parameters) rather than the mocks used to calculate covariance matrices $\boldsymbol{C}$, so all tests in this section do not affect the values of the PCA components (i.e. $\boldsymbol{p}$).}. 

In this section, we still adopt the PCA models trained by $\delta W^{\boldsymbol{\alpha}}_{\Delta s}$ and $W^{\boldsymbol{\alpha}}_{\rm 2d}$ with $\boldsymbol{\alpha} = [-0.3,0.3,1]$, and the $N_c$ is determined by $r_c = 0.995$ and $r_c = 0.95$, respectively. Different from  Sec.\ref{sec:result_PCA}, we select randomly 256 sub-box samples from a total of $8^3$ sub-box samples to calculate the new covariance matrices. 
Then we run MCMC again and compare the results with those in Sec.\ref{sec:results_coarse} and Sec.\ref{sec:result_PCA} in Fig.\ref{fig:MCMC_PCA_covRandom}. We find that, in cases of using PCA, there is only a slight difference ($\lesssim 15\%$ change of specific uncertainty of $\Omega_m$ and $w$) in the constraining powers, implying that this strategy is rather robust on covariance estimation. As for the case of using the coarse binning scheme, the difference of constraining parameters is greater, but still below $25\%$.

\begin{figure*}[!htpb]
    \centering
    \includegraphics[width=0.95 \textwidth]{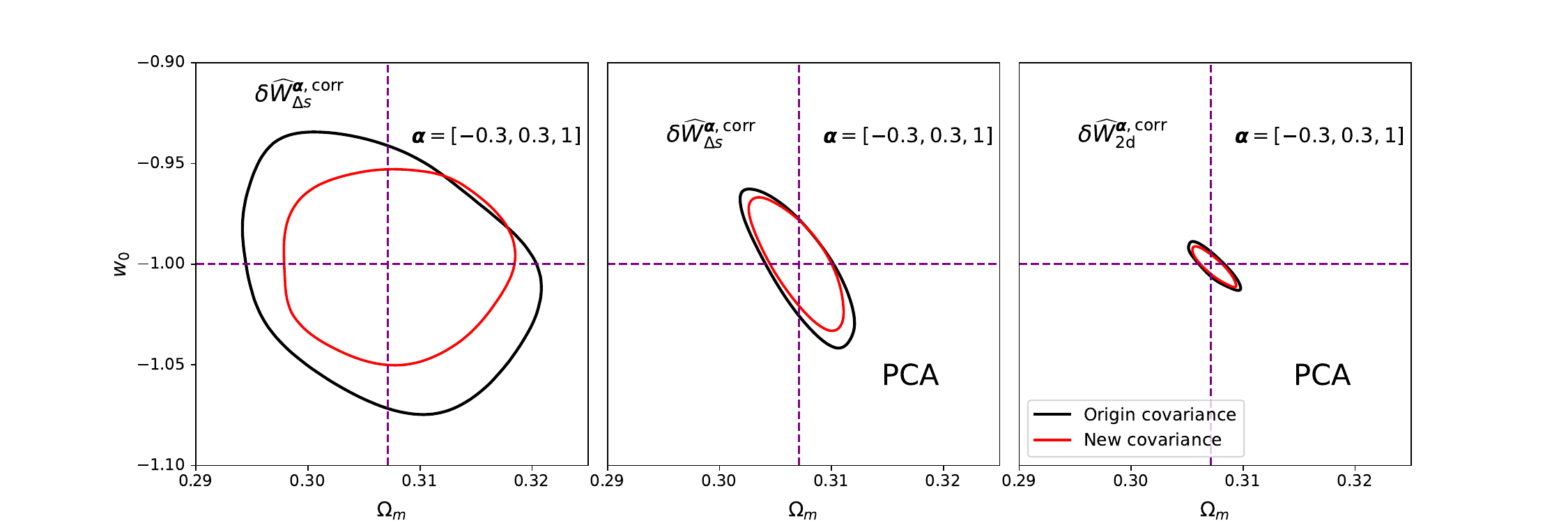}
    \caption{The comparison of $\Omega_m-w$ constraints of combined MCFs with $\boldsymbol{\alpha} = [-0.3,0.3,1]$, between origin and new covariance matrices, the former of which is estimated by entire $8^3$ sub-box samples while the latter of which is estimated by $256$ sub-box samples selected randomly. The left and middle panel shows the constraints of combined MCFs in 1D case with coarse binning method and fine binnings scheme with PCA compression, respectively, while the right panel shows the constraints of combined MCFs in 2D case with PCA. The dashed lines show fiducial values of $\Omega_m$ and $w$.}
    \label{fig:MCMC_PCA_covRandom}
\end{figure*}

\section{A preliminary test of MOPED} \label{sec:appendix_MOPED}

In this section we test the performance of using MOPED rather than PCA to compress the data vectors. MOPED is a compression method which can form linear combinations of the data which contain as much information, in the sense that the Fisher information matrices are identical, if the noise in the data is independent of the parameters; i.e. the method is lossless\citep{MOPED}.

In MOPED, it suggests the measurements vector $\boldsymbol{x}$ includes a signal and noise part, which denotes $\boldsymbol{\mu}$ and $\boldsymbol{n}$, respectively,
\begin{equation}
    \boldsymbol{x} = \boldsymbol{\mu} + \boldsymbol{n}
\end{equation}
In our work, we assume the signal part $\boldsymbol{\mu}$ can be estimated by $\delta \widehat{W}_{\Delta s}^{\boldsymbol{\alpha}, \rm corr}$ with $\boldsymbol{\alpha} = [-0.3, 0.3,1]$ while the noise part can be estimated by the covariance matrix $\boldsymbol{C}$ in Eq.\ref{eq:chi2}.  

Based on the mathematical requirement of maintaining the Fisher information matrix identical, MOPED leads to a simple linear transformation
\begin{equation}
    y \equiv \boldsymbol{b}^{T} \boldsymbol{x}
\end{equation}
\noindent where $\boldsymbol{b}$ is determined by the differential between $\boldsymbol{\mu}$ and cosmological parameters to constrain ($\Omega_m$ and $w$ in this work), along with the inverse covariance matrix $\boldsymbol{C}^{-1}$. The size of $\boldsymbol{b}$ is the same as the cosmological parameters to constrain, therefore, it is fixed as $2$ in this work. It is constructed as 
\begin{equation}
    \begin{aligned}
        \boldsymbol{b}_1 &= \frac{\boldsymbol{C}^{-1} \boldsymbol{\mu}_{,1}}{\sqrt{\boldsymbol{\mu}_{,1} \boldsymbol{C}^{-1} \boldsymbol{\mu}_{,1}}} \\
        \boldsymbol{b}_2 &= \frac{\boldsymbol{C}^{-1}\boldsymbol{\mu}_{,2} - (\boldsymbol{\mu}_{,2}^{T} \boldsymbol{b}_1 ) \boldsymbol{b}_2 }{\sqrt{\boldsymbol{\mu}_{,1} \boldsymbol{C}^{-1} \boldsymbol{\mu}_{,2} - (\boldsymbol{\mu}_{,2}^{T} \boldsymbol{b}_1 )^{2}}}
    \end{aligned}
\end{equation}

\noindent where suffix $1$ and $2$ denote the first and second cosmological parameters to constrain, and the order of them has no effect. Without losing generality, we set the first and second parameters as $\Omega_m$ and $w$ in this work. 

After the transformation, we run MCMC again to obtain the constraints with MOPED. The comparison between PCA and MOPED is in Fig.\ref{fig:MCMC_PCA_MOPED}, without redshift errors and with redshift errors, the form of which is $\sigma_z = 0.002(1+z)$. The strategy to choose $N_c$ in PCA cases is the same as Sec.\ref{sec:result_PCA}. 

The improvement of MOPED is significant, leading to a $\simeq 55\%$ improvement in the constraints of $\Omega_m$ and $w$ in both cases of with or without redshift errors. Besides, the MOPED compression is also capable of avoiding redshift error interference, as the effect of redshift error is slight ($\simeq 5\%$). 

Unfortunately, we fail to apply MOPED to the constraints of 2D cases. The reason is that the MOPED transformation needs to be done after the computation of the covariance matrices, which are rather difficult due to the very large size of statistics vectors. So in this work we still mainly focus on the PCA compression. Nevertheless, MOPED's excellent characteristics merit additional research within our framework. We are going to do further tests on it in the future, for example, to find a good way to combine PCA or other efficient compression methods and MOPED in the constraints to balance data compression and the information lossless level, especially in the 2D cases.

\begin{figure*}[!htpb]
    \centering
    \includegraphics[width=0.95 \textwidth]{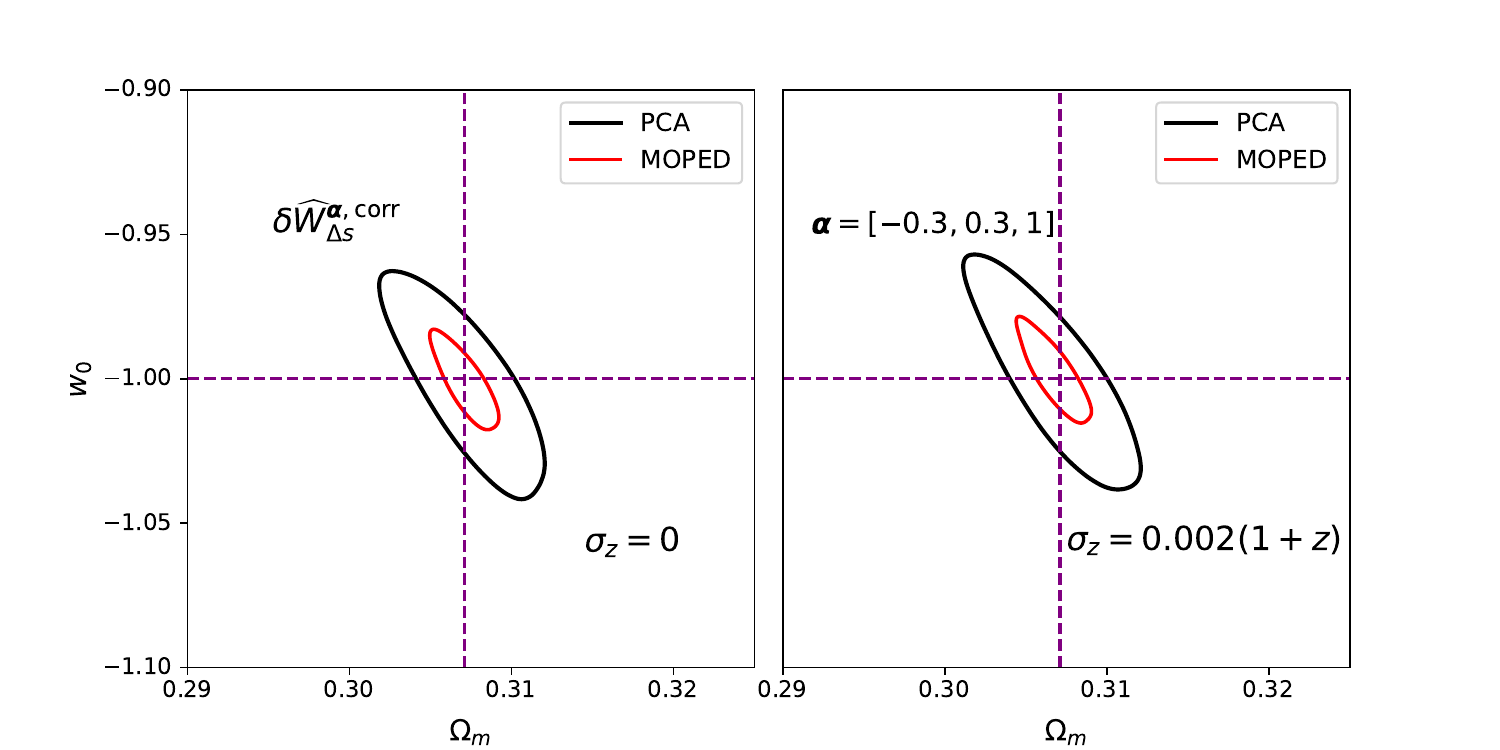}
    \caption{The comparison of $\Omega_m-w$ constraints between PCA and MOPED compression in the cases of 1-dimensional combined MCFs with $\boldsymbol{\alpha} = [-0.3,0.3,1]$, without redshift errors (RE) and with redshift errors, the form of which is $\sigma_z = 0.002(1+z)$. The dashed lines show fiducial values of $\Omega_m$ and $w$.}
    \label{fig:MCMC_PCA_MOPED}
\end{figure*}

\section{A simple extension to a CSST-like survey} \label{sec:appendix_extension}

It would be useful to have a rough estimation of the power of cosmological constraints for a CSST-like survey. In this work, the analysis is done utilizing the BIGMD mock with a length of $2500 \mathrm{Mpc}/h$ and fix the density of halos and subhalos with $\overline{n} = 10^{-3}$ in the two snapshots of $z=$ 1.0 and 0.6069. We extend our results to a CSST-like survey based on the conclusion in ~\citep{LI19}. The authors found that the covariance matrices scale with $N_{\rm gal}$ as
\begin{equation}
    \mathrm{Cov} \propto 1 / N_{\rm gal}
\end{equation}\label{eq:cov_to_N}
\noindent where $N_{\rm gal}$ is the number of galaxies. They found Eq.\ref{eq:cov_to_N} is not a very bad approximation when the changes in the number densities of the samples are not significant.

From ~\citep{Miao_2022}, the number densities in the CSST spectroscopic survey are $5.63 \times 10^{-3}$ and $1.15 \times 10^{-3}$ in the spec-z bins of $0.6 \le z \le 0.9$ and $0.9 \le z \le 1.2$, respectively, which are in the same order of magnitude as the number density of mocks used in our work. The sky coverage of CSST is around 17,500 $\rm deg^2$, and therefore, the ratio of comoving volumes between two CSST spec-z bins and BIGMD are around 0.84 and 0.95, respectively. Based on these facts, Eq. \ref{eq:cov_to_N} tells us that the ratio of variance between CSST and BIGMD is around 0.6. So for CSST, the results of constraints are roughly 30\% tighter than those shown in this work.

We emphasize that the above estimates are very preliminary because 1) the exact redshift error characterization for CSST’s slitless survey is not yet established; 2) in the analysis of CSST data one can adopt many more redshift bins to improve cosmological constraints; 3) the above estimation excludes lower-redshift galaxies ($z<0.6$) from the CSST survey. We plan to conduct a more comprehensive analysis of CSST survey forecasts in future work.

\end{document}